\documentclass[12pt, a4paper]{article}
\renewcommand{\baselinestretch}{1.5}
\usepackage{amsmath}
\usepackage{epsfig}
\usepackage{amssymb}
\usepackage{amscd}
\usepackage{palatino}
\usepackage{graphics}
\usepackage{enumerate}
\usepackage{hhline}
\usepackage{natbib}
\usepackage{float}
\usepackage{chngpage}
\usepackage{rotating}
\usepackage{multirow}
\usepackage{color}
\usepackage{colortbl}
\usepackage[separate]{patty}

\def\max{\mathop{\rm max}\limits}
\usepackage{authblk}

\begin{document}
\bibliographystyle{apsr}
\renewcommand{\baselinestretch}{1}

\title{Ban The Box? Information, Incentives, and Statistical Discrimination 
}

\author{John W. Patty\thanks{Professor of Political Science and Quantitative Theory \& Methods, Emory University.\\Email: \textit{jwpatty@gmail.com}.} \and Elizabeth Maggie Penn\thanks{Professor of Political Science and Quantitative Theory \& Methods, Emory University.\\Email: \textit{elizabeth.m.penn@gmail.com}.}}

\date{\today}

\maketitle

\begin{abstract}

``Banning the Box'' refers to a policy campaign aimed at prohibiting employers from soliciting applicant information that could be used to statistically discriminate against categories of applicants (in particular, those with criminal records). In this article, we examine how the concealing or revealing of informative features about an applicant's identity affects hiring both directly and, in equilibrium, by possibly changing applicants' incentives to invest in human capital. We show that there exist situations in which an employer and an applicant are in agreement about whether to ban the box.  Specifically, depending on the structure of the labor market, banning the box can be (1) Pareto dominant, (2) Pareto dominated, (3) benefit the applicant while harming the employer, or (4) benefit the employer while harming the applicant. Our results have policy implications spanning beyond employment decisions, including the use of credit checks by landlords and standardized tests in college admissions.

\mbox{}\\

\noindent \textbf{Word Count (excluding online Appendix): 11,903.}\\

\end{abstract}

\newpage

\renewcommand{\baselinestretch}{2}
\normalsize
\section{Introduction}

Discrimination is pervasive across political, economic, and social settings, including the markets for housing, credit, and employment.   Eliminating it is a longstanding and vexing policy challenge.  While the term ``discrimination'' itself has a wide array of closely related definitions, in this article we say that discrimination occurs whenever a decision-maker treats one group of applicants differently than another group, simply as a function of their group memberships (\textit{i.e.}, holding all other factors equal).  

  Discrimination can arise from various sources, including a ``taste'' for one group over another (\textit{e.g.}, \cite{Becker71}),  belief-based ``statistical'' discrimination (\textit{e.g.}, \cite{Phelps72}, \cite{Arrow73Discrimination}), and implicit biases in evaluating/choosing individuals (\textit{e.g.}, \cite{BertrandChughMullainathan05}).  A feature common to all three of these sources of discrimination is that the employer must be able to observe (or infer) applicants' group memberships.\footnote{We use hiring as our running example in this article, but the implications are more general in scope.}  Accordingly, eliminating or withholding this information may help forestall discrimination at its root source.\footnote{A related issue (that for reasons of space we do not confront as squarely as we could in this article) is the degree to which employers can, or should, infer sensitive information about applicants from seemingly innocuous co-variates.  Our theory does indicate the importance of this question to the degree that it clearly, if partially, illustrates the situations in which such an incentive would emerge \textit{in equilibrium}.}  
  
  This article examines the theoretical impact of withholding potentially sensitive information that an employer might use to discriminate between applicants.  A central conclusion of our analysis is that withholding such information can have ambiguous welfare effects.  In many cases, the information has no impact on welfare whatsoever, but withholding the information can (1) hurt employers while helping workers, (2) hurt workers while helping employers, (3) hurt both employers and workers, or (4) help both employers and workers.  
  
  Our model of the labor market is very stylized, largely focusing only one dimension of the employment market (specifically, moral hazard).  However, this simplicity implies that the theory indicates some reasons why the policy debate surrounding \textit{how} to reduce discrimination is best not simply thought of as a zero-sum struggle between labor and management.

\subsection{Information, Discrimination, and Incentives}

It is well-known that discrimination can have ``upstream'' consequences in the sense that the expectation of discriminatory practices might differentially affect the incentives of individuals to invest in skills, experiences, and expertise that can help them succeed later in life (\textit{e.g.}, \cite{CoateLoury93}, \cite{LundbergStartz98}).  We build on this body of work by considering how eliminating the information required for direct discrimination affects incentives on both sides of the market, employment outcomes, and overall welfare.  Specifically, we consider how hiding applicants' group memberships from potential employers will affect the employers' willingnesses to hire, which will then in turn determine the applicants' incentives to invest in becoming qualified for the job in question.\footnote{\cite{Eguia17} and \cite{KimLoury19} consider related models in which group membership may be endogenously determined.}

\section{Ban the Box}

A policy proposal that motivates our theory is popularly known as ``Ban the Box,'' or BTB.\footnote{For a comprehensive review of the theoretical and empirical literatures about BTB, see \cite{Raphael20}.}  Such policies have been both adopted voluntarily by some employers, such as Starbucks, Target, and Walmart, and imposed by law in various states and localities.  In practice, BTB policies generally preclude employers from considering a job applicant's criminal history at least initially in the hiring process.\footnote{The point at which such consideration is permissible varies across jurisdictions.  It is common for the prohibition to extend until a conditional offer of employment is made.  A full consideration of the effect of this timing is very interesting.  However, space precludes us from treatment of this issue in this article.}  

\subsection{Banning The Box and Statistical Discrimination}

Discrimination comes in a variety of forms (\textit{e.g.}, \cite{NationalResearchCouncil04}). A classic division of these is between \textit{direct} and \textit{indirect} forms of discrimination.  Direct discrimination occurs when the hiring decision is conditioned upon an applicant's group membership.  Indirect discrimination occurs when the hiring process produces different outcomes for different groups.  Obviously, removing information about an applicant's criminal record removes the simplest route for direct discrimination.\footnote{In our model, it removes the only route for direct discrimination. We say that it removes the ``simplest'' route because, in reality, it is possible that the employer could obtain this information through means beyond the scope of our model.}  However, removing the information may induce the employer to not hire any applicants.  When this occurs in our model,\footnote{Specifically, when the parameters of the model are such that this ``market failure'' occurs in equilibrium without group membership information but there is a positive rate of employment in equilibrium when information about group membership is available.} this behavior represents belief-based statistical discrimination because there is no equilibrium in which the employer can hold accurate beliefs that justify hiring any applicant without knowing the applicant's group membership.  

There is empirical evidence of statistical discrimination occurring when the box is banned (\textit{e.g.}, \cite{DoleacHansen20}).  However, the answer to the central policy question --- namely whether the negative effect of statistical discrimination on all workers under BTB outweighs the positive effect of eliminating disparate treatment of members of the disadvantaged group --- depends on several parameters, including ``the extent to which those with criminal histories benefit from suppressing information, the extent to which those without criminal histories are harmed, and the relative size of these two classes of applicants within the group itself'' (\cite{Raphael20}, p.7).  

It is important to note that, though BTB is frequently discussed regarding consideration of criminal history in the hiring process, the logic identified by our model regarding the ambiguity of BTB's effects on both employment and welfare extends to the effects of including/excluding other types of information, including information about credit history (\textit{e.g.}, \cite{BartikNelson19}, \cite{MaturanaNickersonTruffa20}) in both hiring and other decision-making processes, such as housing.  

\section{Our Model}

The existing literature on discrimination tends to focus on either \textbf{taste-based discrimination} (\textit{e.g.}, \cite{Becker71}), in which an employer prefers workers from one group over another, or \textbf{statistical discrimination} (\textit{e.g.}, \cite{Phelps72}), in which an employer holds different beliefs about a worker's unobserved characteristics based on the worker's group membership.  As displayed in Table \ref{Tab:FormsOfDiscrimination}, our theory pursues the latter path, asking the question of how removing the information required for ``direct'' discrimination (for example, explicitly and knowingly hiring one group at a lower rate than another) can affect investment, employment, and welfare through its potential to incentivize ``indirect'' statistical discrimination in equilibrium. To keep our analysis as compact as possible without distorting the underlying analysis, we rule out wage discrimination and focus solely on employment discrimination in terms of differential standards for employment at a prevailing, common market wage.\footnote{We discuss relaxing the assumption that the employer is a wage taker in Section \ref{Sec:Conclusion}.}

\begin{table}[hbtp]
\centering
\begin{tabular}{|c|c|c|}
    \hline
\multicolumn{1}{|c|}{{\bf Decision-Maker's}} & \multicolumn{2}{c|}{{\bf Origin of Discrimination}} \\
    \cline{2-3}
\multicolumn{1}{|c|}{{\bf Information}}    & \textit{Inherent Group Differences}    & \textit{Group Membership, \textit{per se}} \\
    \hline
\textit{Group Observed}  & \cellcolor[gray]{0.9} Direct Statistical            & Direct Taste-Based \\
    \hline
\textit{Group Inferred}  & \cellcolor[gray]{0.9} Indirect Statistical          & Indirect Taste-Based \\
    \hline
\end{tabular}
\caption{
A Typology of Forms of Discrimination (Our Theory: Gray Cells)
\label{Tab:FormsOfDiscrimination}}
\end{table}

In our theory, workers have correct expectations about employers' approach to hiring when deciding whether to become qualified or not, as in \cite{CoateLoury93}. The two groups of workers are distinguished solely by the probabilities that the members of the two groups have an opportunity to become qualified at all.  We refer to this probability as the group in question's {\bf potential}.  One of the immediate conclusions within our framework is that BTB can have an effect on outcomes only if the groups differ in terms of their potentials and/or the ability for the employer to detect a group members' true qualification.\footnote{\label{Fn:EquilibriumSelection} As we discuss later (footnote \ref{Fn:EquilibriumSelection2}), our analysis assumes that the presence or the absence of the box has an effect on outcomes simply by serving as ``an equiibrium selection mechanism.''}  Because we assume that the employer has equally accurate information about the skills acquired by individuals in both groups, our theory indicates that banning the box can help eliminate discrimination in practice only to the degree that the groups have different likelihoods of having the opportunity to acquire the skills desired by employers.\footnote{This distinguishes our analysis from that of \cite{CoateLoury93}, who assume that the groups have identical potentials.  In addition, our results also provide another justification for the assumption in Coate and Loury's analysis that the employer observes each applicant's group membership.}

We consider a simple situation in which there are two groups, differing in their potentials.  We refer to the group with higher potential as the {\bf advantaged group} and the other group as {\bf disadvantaged}.  In terms of the application to BTB, our analysis and presentation supposes that --- holding all else constant --- an individual who has been convicted of a felony has had a lower probability of having the opportunity to acquire job-relevant skills prior to applying for employment than an otherwise similar applicant who has not been convicted of a felony.

\begin{remark}
\emph{For simplicity, our theory considers only two groups of workers.  In the context of BTB, we personally think of these two groups as ``convicted felons'' and ``all other people,'' but the debate about BTB is in a sense really about a setting with multiple, overlapping groups (\textit{e.g.}, individuals have both racial group membership(s) and felon/non-felon status).  Specifically, much of the debate in the US is about how omitting information about felon status will differentially affect workers of color relative to white workers.  Interpreting our theory's conclusions in the context of this much richer debate requires one to ``step back'' from the model in a sense and, for example, consider the analysis in parallel---one set of parameters for (say) white workers and another set of parameters for workers of color.  Placing these parallel analyses side-by-side will then allow one to consider the impact of BTB on the outcomes experienced by workers from different racial groups as a function of whether information about the second, ``felon status'' group membership is included in the hiring process.}
\end{remark}

Again following \cite{CoateLoury93}, we model the dilemma facing employers and workers as a moral hazard problem: any given worker's investment in qualification is imperfectly observed by the employer.  As mentioned above, we assume that the precision of the employer's noisy signal about any given worker's true qualification is independent of the worker's group membership.  In this setting, our theory identifies the induced preferences regarding BTB for each of the three types of actors (\textit{i.e.}, employers, and workers in each of the two groups).  A few of the more notable results from the baseline model are as follows.
\begin{enumerate}
    \item BTB makes employers better off only if the box's presence induces workers in one or more groups to choose to \textit{not} become qualified in some situations in which they would choose to become qualified if the box were absent.  
    \item BTB can affect outcomes only 
    \begin{enumerate}
        \item when the employer believes that the two groups have substantively different potentials (Corollary \ref{banningNoMatter1}),
        \item when the employer is able to infer differences in qualifications with sufficient precision (Corollary \ref{banningNoMatter2}).
        \end{enumerate}
    \item When the employer's information about qualifications is very precise, BTB
    \begin{enumerate}
        \item helps disadvantaged workers (if the population at large has high potential), or
        \item hurts disadvantaged workers (if the population at large has low potential), but
        \item always hurts the employer (Corollary \ref{Cor:BTBUniformlyInformative}).
    \end{enumerate} 
    \item Finally, and perhaps most surprisingly,
    \begin{enumerate}
  \item    BTB can help employers while hurting workers (Proposition \ref{mixedOpposed}), and 
  
 \item BTB can Pareto dominate the box (Proposition \ref{Pr:BTBParetoEfficientMSE}).
 
\end{enumerate}
\end{enumerate}




\noindent With the basic outline of our theory's key results in hand, we now turn to a brief discussion of some related models of discrimination in political economy.

\subsection{Related Models}

\cite{Becker71} presented the seminal analysis of the economics of \textit{taste-based} discrimination.  \cite{Phelps72} and \cite{Arrow73Discrimination} presented early models of statistical discrimination, including how such discrimination can be self-enforcing. \cite{CoateLoury93} extended this line of inquiry to consider how discrimination affects the incentive to invest in human capital and, relatedly, whether affirmative action policies might break this self-enforcing nature. \cite{MoroNorman04} combine the \cite{Arrow73Discrimination} and \cite{CoateLoury93} models within a task-assignment context.  \cite{Fryer07} considers the interaction of discrimination at the hiring stage and in subsequent promotion decisions.  \cite{Bjerk08} extends the study of dynamic, statistical discrimination by considering how differences in an employer's informational precision early in an applicant's career might affect the promotion path.

Of these, the model developed by \cite{CoateLoury93} is the most closely related to ours,\footnote{Indeed, it was part of the inspiration for this research.} so it is useful to consider the distinction between our model and theirs.  In Coate and Loury's model, the two groups of applicants are identical from an ex ante perspective.  In our model, the two groups are different in the sense that one group is more likely to ``be able to afford to become qualified'' than the other.  It is important to note that this assumption is conservative relative to theirs in the sense that it offers an initial justification for the employer including ``the box'' to distinguish between the two groups.  This is ``conservative'' because our \textit{main conclusion is that there will still exist situations in which the employer strictly benefits from banning the box}.  

In Coate and Loury's model (as in Arrow's), discrimination can emerge in equilibrium in spite of the fact that the groups are identical in ex ante terms.  Discrimination in such settings results from equilibrium multiplicity in the labor market and occurs when each applicant's group membership essentially serves as an equilibrium selection device.  Accordingly, within their setting, the impact of banning the box would depend upon which equilibrium would be played in the absence of the box.  As in Coate and Loury's model, our model typically has multiple equilibria.  However, our arguments do not leverage this multiplicity: we focus throughout only on the (generically unique) Pareto efficient equilibrium.  Accordingly, our comparison of labor markets with and without the box presumes that the box plays no role in equilibrium selection.\footnote{\label{Fn:EquilibriumSelection2} This equilibrium selection issue (alluded to in fn.~\ref{Fn:EquilibriumSelection}, above), and how it connects a few seemingly disparate models of statistical discrimination, are each also addressed briefly in \cite{PattyPenn21PhilCompass}.}

\section{A Moral Hazard Model\label{Sec:SingleGroup}}

Our theory is based on a two player game involving a {\bf worker}, $W$, and an {\bf employer}, $E$.  In order to better illustrate the incentives of this baseline model, we consider first a setting with only one group of workers.  

\paragraph{The Worker's Information \& Potential.} The worker has a (privately observed) binary real-valued \textit{type}, $c \in C \equiv \{c_{L},c_{H}\}$, with $0<c_{L}<c_{H}$. The type determines the cost of becoming qualified ($q=1$) relative to remaining unqualified ($q=0$) and is distributed as follows:
\begin{eqnarray*}
\Pr[c=c_{L}] & = & p,\\
\Pr[c=c_{H}] & = & 1-p.
\end{eqnarray*}
As mentioned above in the introduction, we refer to the parameter $p$ as the \textbf{potential} of the worker's group.  This is because, in the cases of interest in our analysis (Assumption \ref{As:WorkerCosts}, below), $p$ is the maximum ex ante probability that a worker in that group might actually be qualified in equilibrium.

\paragraph{The Worker's Choices.}  The worker first observes his or her \textbf{cost of qualification}, $c \in \{c_L,c_H\}$ (with $0<c_L<c_H$), and then chooses whether to become qualified (denoted by $q=1$) or not (denoted by $q=0$).  If the worker chooses to become qualified, he or she incurs a net cost of $c$.

\paragraph{The Employer's Information.} The worker's qualification (\textit{i.e.}, $W$'s choice of $q$) is not directly observed by the employer.  Rather, an informative --- but noisy --- signal of his or her choice, denoted by $\theta \in \Theta \equiv \{1,2,3\}$, is generated as follows:
\begin{equation}
\label{Eq:TestingStructure}
\begin{array}{|c|c|c|}
 \hline
 					& q=0 & q=1 \\
\hline
\Pr[ \theta=1 \mid q] 	& \phi_{0} & 0 \\
\Pr[ \theta=2 \mid q] 	& 1-\phi_{0} & 1-\phi_{1} \\
\Pr[ \theta=3 \mid q] 	& 0 & \phi_{1} \\
\hline
\end{array}
\end{equation}
Note that if $E$ observes either $\theta=1$ or $\theta=3$, then the test result reveals the qualification of the worker, $q$, with certainty.   On the other hand, a test result of $\theta=2$ is a ``{\bf garbled test result}'' that can potentially be sent by both qualified and unqualified types.
Accordingly, for each qualification choice, $q \in \{0,1\}$,  $\phi_{q} \in (0,1)$ is the conditional probability that $\theta$ is ``correct'' in the sense of revealing $q$.  We refer to the conditional distribution of $\theta$ described in \eqref{Eq:TestingStructure} as a \textit{test} of qualification, so that $\theta$ represents the \textit{outcome} of the worker's test.    

\paragraph{The Employer's Choices.}  After (1) the worker's cost of qualification, $c$, is realized by the worker, (2) the worker chooses his or her qualification, $q$, and (3) conditional on $W$'s choice of $q$, the test result $\theta$ is realized and observed by the employer, the employer then finally chooses whether to hire $W$ (denoted by $h=1$) or not (denoted by $h=0)$.  

\paragraph{Sequence of Play.}  Summarizing the description above, our model's decision sequence is as follows: 
\begin{enumerate}
\item The worker observes $c$,
\item The worker chooses $q \in \{0,1\}$,
\item The employer observes $\theta$,
\item The employer chooses $h\in \{0,1\}$,
\item The process concludes and players receive their payoffs.
\end{enumerate}

\paragraph{Payoffs.}  The players' payoffs, given $c,q$, and $h$, are as follows:
\begin{equation}
\label{Eq:Payoffs}
\begin{array}{rcl}
u_{W}(q,h\mid c) & = &  w h - c q,\\
u_{E}(h\mid q,\theta) & = & (Bq-w)h,
\end{array}
\end{equation}
where $w>0$ and $B>w$ are exogenous parameters that are assumed to be common knowledge throughout. The parameter $w$ represents the \textbf{wage} paid by $E$ to $W$ if $E$ hires $W$, and $B>w$ represents $E$'s \textbf{benefit} from hiring ($h=1$) a qualified worker ($q=1$). Finally, as noted earlier, $c \in \{c_L,c_H\}$ captures $W$'s cost of obtaining qualification.  

\paragraph{Strategies.} A (possibly mixed) qualification strategy for $W$ is a mapping $\chi: C \rightarrow [0,1]$, where $\chi(c)\equiv \Pr[q=1\mid c]$ denotes the probability that the worker chooses $q=1$, given his or her cost, $c$.  Similarly, a (possibly mixed) hiring strategy for the employer is a mapping $\eta: \Theta \rightarrow [0,1]$, where $\eta(\theta) \equiv \Pr[h=1\mid \theta]$ denotes the probability $E$ hires $W$, for each $\theta \in \Theta$.  We refer to the employer's hiring strategy, $\eta$, as \textbf{aggressive} when $\eta(2)=1$, \textbf{conservative} when $\eta(2)=0$, and \textbf{mixed} when $\eta(2)\in(0,1)$.

\paragraph{Beliefs.} The employer's beliefs about $q$, given $\theta$, are denoted by $\mu(\theta) \equiv \Pr[q=1\mid \theta]$.  Our equilibrium concept of choice, sequential equilibrium, will require that these beliefs be correct.  We now turn to the analysis of the model.

\subsection{Equilibrium Analysis}

Our equilibrium concept is sequential equilibrium (\cite{KrepsWilson82}), a refinement of perfect Bayesian equilibrium.  Sequential equilibria are typically more complicated to verify than perfect Bayesian equilibria, but have the benefit of ruling out some perfect Bayesian equilibria in which $E$ holds ``unreasonable off the path beliefs.''  In our setting, this refinement is particularly useful because it rules out an otherwise ubiquitous perfect Bayesian ``pooling'' equilibrium in which the employer never hires workers (even after observing $\theta=3$) and workers never become qualified.  This is not a sequential equilibrium: in any sequential equilibrium, $E$'s beliefs about $q$ must satisfy the following:\footnote{To see this, consider any sequence of ``fully mixed'' strategies by the worker, $\{\chi_\tau\}_{\tau=1}^{\infty}$ with $\chi_\tau(c) \in (0,1)$ for both $c \in \{c_{L},c_{H}\}$, and consider the sequence of beliefs, $\{\mu^*_\tau\}_{\tau=1}^{\infty}$ such that $\mu^*_\tau$ is consistent with $\chi_\tau$ and Bayes's rule for each $\tau \in \{1,2,\ldots,\}$.  This is uniquely defined for each $\tau \in \{1,2,\ldots\}$ and satisfies the following: $\mu^*_{\tau}=1$ for all $\tau \in \{1,2,\ldots\}$.}
\begin{eqnarray*}
\mu(3) & = & 1.
\end{eqnarray*}
With this in hand, we can simplify notation and write $E$'s beliefs simply as $\mu \equiv \mu(2) \in [0,1]$.\footnote{The structure of the payoffs in \eqref{Eq:Payoffs}, along with the assumption that $\phi_0>0$, imply that $\Pr[\theta=1]>0$ in any Bayes Nash equilibrium of this model, so that Bayes's rule implies that $\mu(1)=0$ in any Bayes Nash equilibrium.} $E$'s beliefs, $\mu$, are consistent with $W$'s strategy, $\chi$, if $\mu$ satisfies the following:
\begin{equation}
\label{Eq:SingleGroupConsistency}
\mu = \frac{(1-\phi_{1}) (p\chi(c_{L})+(1-p)\chi(c_{H}))}{(1-\phi_{1}) (p\chi(c_{L})+(1-p)\chi(c_{H}))+(1-\phi_{0}) (p(1-\chi(c_{L}))+(1-p)(1-\chi(c_{H})))}.
\end{equation}

\paragraph{Equilibrium Hiring.} The sequentially rational hiring strategy for $E$, given $\mu$, is essentially defined by the following:
\begin{equation}
\label{Eq:SingleGroupSequentialRationality}
\eta(\theta\mid \mu) = \begin{cases}
0 & \text{ if } \theta=1,\\
0 & \text{ if } \theta=2 \text{ and } \mu<\frac{w}{B},\\
1  & \text{ if } \theta=2 \text{ and } \mu> \frac{w}{B},\\
1 & \text{ if } \theta=3,
\end{cases}
\end{equation}
and any hiring probability is sequentially rational conditional upon $\theta=2$ and $\mu=\frac{w}{B}$.  With this in hand, we will write $E$'s strategy simply as $\eta\equiv \Pr[h=1\mid \theta=2]$.

\paragraph{Equilibrium Qualification.}  Turning to the worker, first note that if $w<c_{L}$, then $q=1$ is strictly dominated for $W$, so that $\chi(c_L)=0$ and $\eta=0$ in any equilibrium.  On the other hand, when $c_{H}<w$, there may exist equilibria in which all workers obtain qualification with probability 1 (\textit{i.e.}, $\chi(c_{L})=\chi(c_{H})=1$).  While these equilibria are interesting in their own right, they do not accurately reflect the role we intend for the parameter $p$ to play in the model --- an upper bound on the probability that $q=1$ (\textit{i.e.}, the maximum ``equilibrium potential'' of the worker's group). Accordingly, we assume throughout that  $c_{L}<w<c_{H}$, so that $q=1$ is strictly dominated if $c=c_{H}$ but not strictly dominated when $c=c_{L}$ (this does not imply that $q=1$ is a best response for the worker when $c=c_{L}$). 

\begin{assumption}
\label{As:WorkerCosts}
Qualification is strictly dominated for $W$ conditional on $c=c_H$ and costly, but not strictly dominated, conditional on $c=c_L$:
\[
0 < c_L < w < c_H.
\]
\end{assumption}
With Assumption \ref{As:WorkerCosts} in hand, we simplify notation and write $W$'s strategy simply as $\chi\equiv \Pr[q=1\mid c=c_{L}]$ (\textit{i.e.}, $\chi(c_H)=0$ in all equilibria). We begin with $E$'s sequentially rational hiring decision conditional on $\chi$ and $\theta=2$.  $E$ is willing to hire (i.e. to set $\eta>0$) only if 

\begin{equation}
\label{Eq:Hire2OneGroup}
\frac{(1-\phi_{1}) p \chi}{(1-\phi_{1}) p \chi+(1-\phi_{0}) (1-p+p(1-\chi))}\geq \frac{w}{B}.
\end{equation}  

\ \\  Similarly, conditional on $c=c_L$ and the strategy $\eta$ by $E$, it is incentive compatible for $W$ to play strategy $\chi>0$ only if 
\begin{equation}
\label{getQ}
w\geq \frac{c_L}{\phi_1+\eta(\phi_0-\phi_1)}.
\end{equation}

\ \\ Equations \ref{Eq:Hire2OneGroup} and \ref{getQ} give us two cases to consider, distinguished by the employer's sequentially rational decision conditional on $\theta=2$ when $E$ believes $\chi=1$.  If
\begin{equation}
\frac{(1-\phi_{1}) p}{(1-\phi_{1}) p+(1-\phi_{0}) (1-p)}\geq \frac{w}{B},
\end{equation}
then $E$ is willing to hire upon observing $\theta=2$ if he or she believes that $\chi=1$. Accordingly, in this case there exists an equilibrium with $\chi=1$ if and only if 
\begin{eqnarray}
w & \geq & \frac{c_{L}}{\phi_{0}}. \label{Eq:ICHire2}
\end{eqnarray}
On the other hand, if
\[
\frac{(1-\phi_{1}) p}{(1-\phi_{1}) p+(1-\phi_{0}) (1-p)}< \frac{w}{B},
\]
then $E$ is unwilling to hire conditional on $\theta=2$ regardless of $\chi$. In this case there exists an equilibrium with $\chi=1$ if and only if $w$ is sufficiently high and/or $\theta=3$ is sufficiently likely, conditional on $q=1$:
\begin{eqnarray}
w & \geq & \frac{c_{L}}{\phi_{1}}. \label{Eq:ICHire3}
\end{eqnarray}
Finally, it may be the case that the Pareto efficient equilibrium is a mixed strategy equilibrium, with $W$ using a non-degenerate mixed strategy conditional upon $c=c_L$, and $E$ using a non-degenerate mixed strategy conditional upon $\theta=2$. In this case Equations \ref{Eq:Hire2OneGroup} and \ref{getQ} must hold with equality, implying

\begin{enumerate}\label{Pg:MSERequirements}
    \item \textit{The test is more accurate conditional on being qualified than not}: $\phi_0<\phi_1$, 
    \item \textit{The wage is sufficiently high to sustain positive qualification}: $w\geq \frac{c_L}{\phi_1}$, and
    \item \textit{The players' equilibrium strategies are described by the following:}
        \begin{eqnarray}
        \eta_M \equiv \eta_M(w,\phi,c_L) & = & \frac{w \phi_1-c_L}{w(\phi_1-\phi_0)},\label{Eq:mixedStrategiesEta}\\
        \chi_M(p) \equiv \chi_M(p,B,w,\phi,c_L) & = & \frac{w(1-\phi_0)}{p(B(1-\phi_1)+w(\phi_1-\phi_0))}. \label{Eq:mixedStrategiesChi}
    \end{eqnarray}
\end{enumerate}

We now define the employer's \textbf{hiring threshold}, denoted by $p_E^*$, as the probability of qualification that makes $E$ indifferent about hiring $W$ after observing a garbled test result ($\theta=2$) conditional on $W$ becoming qualified if and only if $W$'s cost of qualification is $c=c_L$ (\textit{i.e.}, $\chi=1$):\footnote{Note that term $p_E^*$ defined in \eqref{Eq:OverlineP} is simply a rearrangement of Equation \eqref{Eq:Hire2OneGroup}.}
\begin{equation}
\label{Eq:OverlineP}
p_E^*\equiv \frac{w(1-\phi_0)}{B(1-\phi_1)+w(\phi_1-\phi_0)}.
\end{equation}
 Putting Equations \eqref{Eq:Hire2OneGroup}, \eqref{Eq:mixedStrategiesEta}, \& \eqref{Eq:mixedStrategiesChi} together, we can characterize six equilibrium regions, depending on $w$, $\phi_0$, $\phi_1$, and $c_L$.  In order to better characterize these regions, we first describe the types of equilibria that can emerge in our framework.

\paragraph{Types of Equilibria.}  In terms of the worker's strategy, $\chi$, our model admits three qualitative types of equilibria:
\begin{itemize}
    \item In a {\bf full qualification equilibrium} (\textit{FQE}), all low-cost workers get qualified: $\chi=1$.  
    \item In a {\bf zero qualification equilibrium} (\textit{ZQE}), no workers get qualified: $\chi=0$.  
    \item In a {\bf mixed strategy equilibrium} (\textit{MSE}), some low-cost workers get qualified and some don't: $\chi \in (0,1)$.\footnote{Note that the full qualification equilibrium and zero qualification equilibria, when they exist, are otherwise independent of the parameters of the model.  The mixed strategy equilibrium, when it exists, on the other hand, is sensitive to the exact values of these parameters.}
\end{itemize}  
With the three classes of equilibria in hand, the following proposition characterizes all equilibria.  It also demonstrates that, when multiple equilibria exist, the worker and employer share the same preferences over these equilibria.  

\begin{proposition}
\label{Pr:EquilibriumCharacterization}
Table \ref{Tab:SingleGroupEquilibria} characterizes all equilibria of the model. When multiple equilibria exist, they are strictly Pareto ranked as follows: the FQE dominates the MSE, which dominates the ZQE.
\end{proposition}
\begin{table}[hbtp]
\centering
\begin{minipage}[]{5in}
\[
\begin{array}{|c|c|}
\multicolumn{2}{c}{\text{Equilibria when }p> p_E^*}\\ \hline 
\text{Parameters (} c_L,w,\phi_0,\phi_1\text{)} & \text{Equilibria} \\ 
\hline
 \phi_0>\frac{c_L}{w}>\phi_1 & 
\begin{array}{c}
\text{FQE with }\chi^*=1, \eta^*=1,,\\
\text{MSE with }\chi^*=\chi_M(p), \eta^*=\eta_M,\\
\text{ZQE with }\chi^*=0, \eta^*=0
\end{array}
\\ \hline
\phi_0>\frac{c_L}{w}\text{ and }\phi_1>\frac{c_L}{w} & \text{FQE with }\chi^*=1, \eta^*=1 \\\hline
 \phi_1>\frac{c_L}{w}>\phi_0 & \text{MSE with }\chi^*=\chi_M(p), \eta^*=\eta_M,\\ \hline
\frac{c_L}{w}>\phi_0\text{ and }\frac{c_L}{w}>\phi_1 & \text{ZQE with }\chi^*=0, \eta^*=0 \\ \hline
\multicolumn{2}{c}{}\\
\multicolumn{2}{c}{\text{Equilibria when }p< p_E^*}\\ \hline 
\text{Parameters (} c_L,w,\phi_0,\phi_1\text{)} & \text{Equilibria} \\ 
\hline
\frac{c_L}{w}>\phi_1 & \text{ZQE with }\chi^*=0, \eta^*=0  \\ \hline
\phi_1>\frac{c_L}{w} & \text{FQE with }\chi^*=1, \eta^*=0 \\\hline
\end{array}
\] 
\end{minipage}
\caption{Equilibria of the Single-Group Case\label{Tab:SingleGroupEquilibria}}
\end{table}
\begin{proof}
Proofs of all numbered results other than corollaries are located in Appendix \ref{Sec:Proofs}.
\end{proof}



\begin{figure}[hbtp]
\centering
\framebox{
\epsfig{file=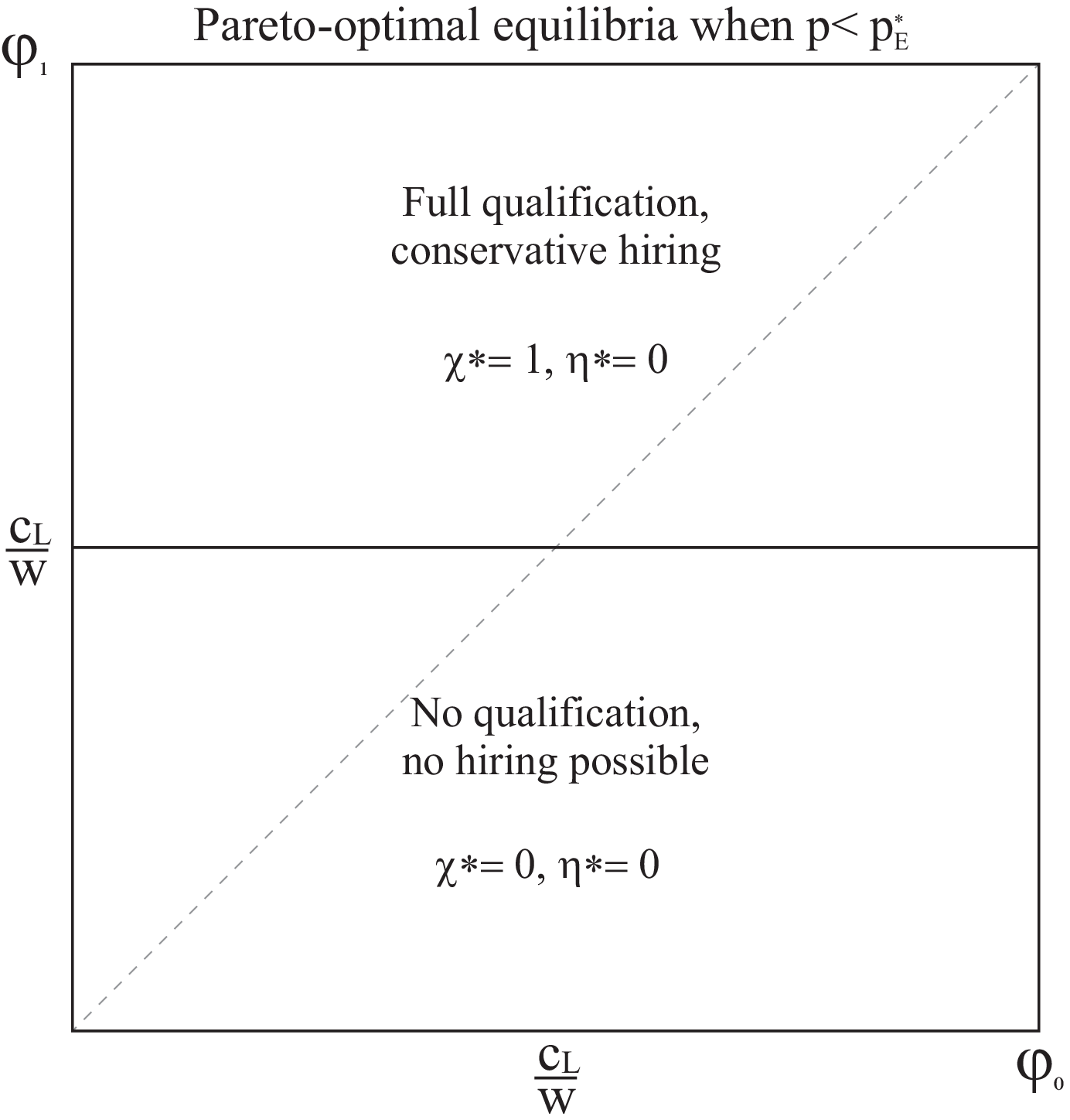, scale=.55}

\epsfig{file=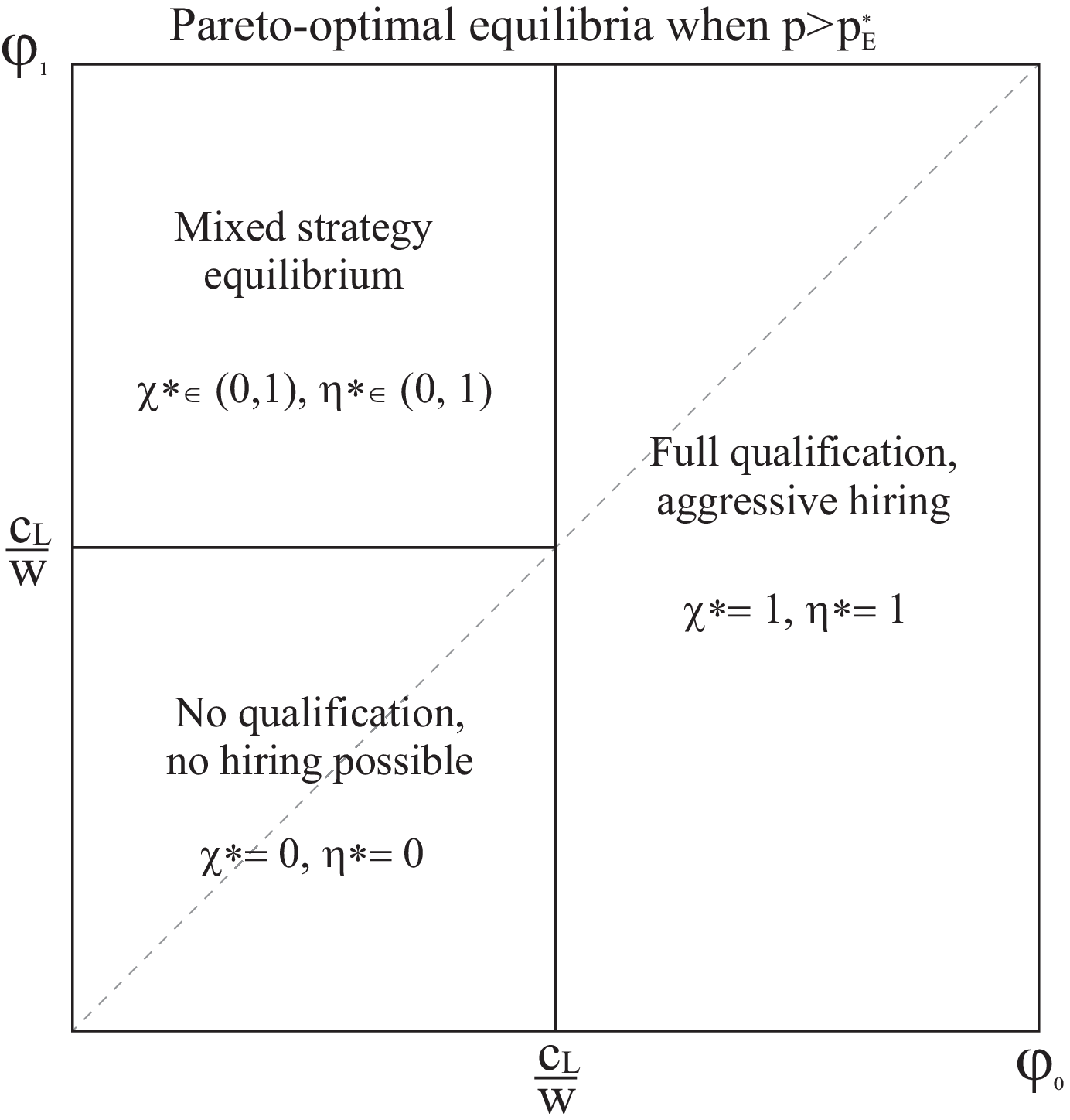, scale=.55}
}
\caption{Pareto-Optimal Equilibria: Single-Group Case \label{Fig:ParetoOptimalEquilibria}}
\end{figure}

Table \ref{Tab:SingleGroupEquilibria} is illustrated in Figure \ref{Fig:ParetoOptimalEquilibria}, which displays the (Pareto efficient) equilibrium regions with respect to the testing technology, $\phi=(\phi_0,\phi_1)$, and the group's potential, $p$, for a given, arbitrary pair of ``low cost level,'' $c_L$, and wage, $w$. The principal point of the single-group analysis is to establish a baseline for examining the effects of labor market heterogeneity and differential information on qualification, employment, and welfare. 

\label{Pg:EqualPhis} Note that, in both panes of Figure \ref{Fig:ParetoOptimalEquilibria},  the 45 degree dashed line represents the continuum of situations in which $\phi_0=\phi_1$ and, as intuition would suggest, there are exactly two possible Pareto optimal equilibria in these cases: when $\phi_0=\phi_1$ is close enough to 0, then the unique equilibrium is a ZQE in which nobody gets qualified and nobody gets hired, because the moral hazard problem is ``too severe'' to sustain credible hiring in equilibrium, and otherwise the unique sequential equilibrium is an FQE in which all workers get qualified if and only if $c=c_L$.  In such cases, the employer's hiring strategy is aggressive if potential ($p$) is sufficiently high and conservative otherwise.

With the equilibrium analysis of the single-group case in hand, we now extend the model to allow for two groups of workers, one of which has greater potential than the other. 

\section{Market Heterogeneity: Two Groups \label{Sec:TwoGroups}}

In this section we maintain the basic structure of the model analyzed above in Section \ref{Sec:SingleGroup} while allowing workers to come from two different groups.  Formally, the worker, $W$, now has a two dimensional {\bf type}, $t = (g,c) \in T \equiv \{1,2\}\times\{c_{L},c_{H}\}$, with $0<c_{L}<c_{H}$, where $g\in \{1,2\}$ denotes the worker's {\bf group}. The worker's type, $t \in T$, is assumed to be distributed as follows:
\begin{eqnarray*}
\Pr[g=1] & = & \gamma,\\
\Pr[c=c_{L}\mid g] & = & p_g,\\
\Pr[c=c_{H}\mid g ] & = & 1-p_{g},
\end{eqnarray*}
with $\gamma \in (0,1)$ representing the proportion of workers who are members of group 1, and $1>p_{1}\geq p_2>0$ representing the potentials of groups 1 and 2, respectively.  

As in the single-group case analyzed above in Section \ref{Sec:SingleGroup}, $c$ represents the cost to $W$ of obtaining qualification.  After observing his or her type, $t=(g,c)$, $W$ chooses $q \in \{0,1\}$, where $q=1$ represents a decision to become qualified.  $E$ then observes both $W$'s group, $g \in \{1,2\}$, and his or her test result, $\theta \in \{1,2,3\}$, distributed conditional on $q$ as described in the single-group case analysis (Equation \eqref{Eq:TestingStructure}).\footnote{As in that section, we assume that both $\phi_q\in (0,1)$ are common knowledge.  Note that this implies that the testing technology is equally informative about $q$ conditional on true qualification, $q$, for workers from both groups.  }  After observing the worker's group membership and test results, $(g,\theta)$, $E$ again chooses to hire $W$ ($h=1$) or not ($h=0$), the game concludes, and the players' payoffs are as defined in \eqref{Eq:Payoffs} for the single-group case.

\subsection{Interpreting Our Model with Respect to Racial Discrimination in the United States}

We consider situations with two groups for reasons of clarity and tractability.  Much of the discussion of the impacts of BTB in the United States are, however, conditional on there being at least four groups: each individual has (1) a racial group membership and (2) a felon/non-felon status. Thus, our welfare analysis is best interpreted as being ``about the impacts of BTB \textit{conditional on racial group membership}.''  Space precludes consideration of a fuller model, but it is a small step from our welfare analysis to a more holistic model with more than two (and potentially overlapping) groups.  With this framing in mind, we now proceed to analyze the equilibrium effects of the box, beginning with the case in which the box is present.

\subsection{Equilibrium Analysis ``With the Box''}  

The two-group model represents the situation facing the worker and employer when the employer can observe the worker's group and condition his or her hiring decision on it.  In other words, $E$ can directly observe $g$, and (importantly) $W$ knows that $E$ can observe $g$.  Formally, the employers' set of information sets (which was $\Theta$ in the single group case) in the two group case \textit{when the box is present} is:
\[
\mathcal{I} = \{1,2\} \times \Theta = \{1,2\} \times \{1,2,3\}.
\]
Equilibrium analysis when the box is present involves simply applying the analysis of the single-group case in Section \ref{Sec:SingleGroup} to each group separately.  Accordingly, we omit a fuller recounting of this analysis and instead turn to consider the effects of ``banning the box,'' or removing the employer's ability to directly condition his or her hiring decision on the worker's group membership.

\section{Banning The Box \label{Sec:BanningTheBox}}

When the box is banned, $E$ cannot condition his hiring decision on $g$. We represent this formally by modifying the game form analyzed above such that the set of information sets for the employer is
\[
\mathcal{I} = \Theta = \{1,2,3\}.
\]
For notational simplicity, we will denote the employer's hiring strategy when the box is banned by $\eta(\emptyset)\equiv \Pr[h=1\mid \theta=2]$ (so as to distinguish it from the single group case analyzed at the outset).\footnote{Note that, in equilibrium when the box is banned, $E$ will learn something about $W$'s group from $\theta=1$ or $\theta=3$.  However, because we have assumed that $E$ does not have a taste for discrimination ($E$ cares only about $q$, not $g$ \textit{per se}), this is irrelevant for our purposes in this article.}   A key point of the analysis is that removing the box has ambiguous welfare effects.  We we are also able to identify some key determinants of the direction, and size, of this effect.  We denote the unconditional probability of an individual having low costs of qualification by $\overline{p}$, which we refer to as the {\bf population potential}:
\[
\overline{p} \equiv \gamma p_1+(1-\gamma)p_2.
\]
We will refer to the population potential as {\bf high} when $\overline{p}\geq p_E^*$ and {\bf low} otherwise.

To describe our results, we label regions of the parameter space.  Table \ref{Tab:GroupPotentials} describes the groups' potentials relative to the employer's hiring threshold, $p_E^*$.\footnote{Note that the case of $p_1=p_2$ is omitted from Table \ref{Tab:GroupPotentials}.  This case is equivalent to the single group case and BTB has no effect on equilibrium behavior.}
\begin{table}[hbtp]
    \centering
    \begin{tabular}{|c|c|}
    \hline
    \textsc{Label} & \textsc{Parameters}\\
    \hline
    {\bf Uniformly high potentials} & $p_1>p_2\geq p_E^*$ \\
    \hline
    {\bf Uniformly low potentials} &  $p_E^*>p_1>p_2$ \\
    \hline
    {\bf Statistically distinct potentials} &  $p_1>p_E^*>p_2$\\
    \hline
    \end{tabular}
    \caption{Typology of Group Potentials
    \label{Tab:GroupPotentials}}
\end{table}

In addition to identifying the importance of group potentials for determining the effect of BTB, our model also illuminates four different types of testing structures.  These are described in Table \ref{Tab:TestStructures}.\footnote{Note that the case of $\phi_0=\phi_1$ is omitted from Table \ref{Tab:TestStructures}.  This case is discussed above on page \pageref{Pg:EqualPhis}.}
\begin{table}[hbtp]
    \centering
    \begin{tabular}{|c|c|}
    \hline
    \textsc{Label} & \textsc{Parameters}\\
    \hline
    {\bf Uniformly informative test} & $\min[\phi_0,\phi_1]>\frac{c_L}{w}$ \\
    \hline
    {\bf Uninformative test} &  $\max[\phi_0,\phi_1]<\frac{c_L}{w}$ \\
    \hline
    {\bf Positively informative test} &  $\phi_1\geq \frac{c_L}{w}>\phi_0$ \\
    \hline
    {\bf Negatively informative test} &  $\phi_0\geq \frac{c_L}{w}>\phi_1$\\
    \hline
    \end{tabular}
    \caption{Typology of Testing Structures \label{Tab:TestStructures}}
\end{table}


With this terminology in hand, we now discuss the impact of the box by working through four qualitative cases, beginning with situations in which the box has no effect on outcomes, moving to situations in which the box's presence affects only the employer's equilibrium hiring behavior, and then concluding with the effects of the box for positively and negatively informative tests, respectively.  Proposition \ref{Pr:EquilibriumCharacterization} provides a road map for our analysis of the equilibrium effects of BTB.  

\subsection{Situations In Which the Box Has No Effect \label{Sec:BoxHasNoEffect}} 

We begin by identifying settings in which the box has no effect on equilibrium behavior.  

\paragraph{Statistically Non-Distinct Group Potentials.}  The fundamental factor in determining whether the box can have an effect is the structure of the employer's potential beliefs in equilibrium, which revolves around the employer's hiring threshold, $p_E^*$ (which is independent of $W$'s group).  Because $\overline{p}$ is a convex combination of $p_1$ and $p_2$, both groups' potentials being greater than $p_E^*$ implies that $\overline{p}$ is also greater than $p_E^*$.  In this case, $E$ will use the same hiring strategy for each group if group identity is observed, and $E$ will also use this same strategy in the event that group identity is not observed ($g=\emptyset$).  Consequently, if both groups have high potentials ($p_1>p_2>p_E^*$), then BTB can have no effect on equilibrium behavior when comparing the Pareto optimal equilibrium in each case.\footnote{As mentioned above (footnote \ref{Fn:EquilibriumSelection} on page \pageref{Fn:EquilibriumSelection}), we are focusing on Pareto efficient equilibria throughout so that the presence or absence of the box does not have an effect on outcomes merely as an equilibrium selection device.}  The same logic follows if both groups have low potential ($p_E^*>p_1>p_2$): in this case, the employer will always use a conservative hiring strategy in equilibrium, regardless of whether the box is present or not.   This leads to the following corollary of Proposition \ref{Pr:EquilibriumCharacterization}, which we present without proof.

\begin{corollary}
\label{banningNoMatter1}
Banning the box can affect Pareto efficient equilibrium behavior and/or welfare only if the groups have statistically distinct potentials: $p_1\geq p_E^* >p_2$.
\end{corollary}  

The following remark is separated out in order to clarify the relationship between this model and other theoretical analyses of discrimination.
\begin{remark}
\textit{Note that, as we do throughout, Corollary \ref{banningNoMatter1} restricts our comparisons to Pareto efficient equilibria.  This focus separates our analysis from that provided by \cite{CoateLoury93} (and many other models of Arrovian statistical discrimination), because in that model, the causal mechanism for discrimination operates through the role of a worker's group membership as an equilibrium selection device.}
\end{remark}

\paragraph{Uninformative Testing Structures.}  Corollary \ref{banningNoMatter1} identifies only a necessary condition for BTB to have an effect on equilibrium behavior.  It is not sufficient --- Figure \ref{Fig:ParetoOptimalEquilibria} also depicts situations in which the groups have statistically distinct potentials, but BTB still has no effect on equilibrium behavior.  This occurs when both $\phi_1$ and $\phi_0$ are low.  In this case the test result is so noisy that it is not in any worker's interest to invest in qualification.  The most straightforward example of this scenario would be when $\phi_1$ and $\phi_0$ both approach zero.  In the limit, every applicant would receive a test score of 2 regardless of qualification status, and no applicant would choose to become qualified.  Again, this is summarized in the following corollary to Proposition \ref{Pr:EquilibriumCharacterization}, which is also presented without proof.

\begin{corollary}\label{banningNoMatter2} When the test is uninformative ($\max[\phi_0,\phi_1]<\frac{c_L}{w}$), BTB has no effect on equilibrium behavior or welfare.
\end{corollary}

Corollaries \ref{banningNoMatter1} and \ref{banningNoMatter2} separately indicate the theoretical limits of BTB as a policy tool for ameliorating discrimination in hiring and, more fundamentally, illustrate the ``informational foundations'' of BTB's impact (or lack thereof) on equilibrium qualification and hiring.  Corollary \ref{banningNoMatter1} highlights that BTB can have an impact on statistical discrimination only if the employer's beliefs about the two groups are statistically distinct, implying that the employer would treat workers from the two groups differently conditional on a garbled test result even if the employer believes that workers from both groups were obtaining qualification whenever qualification is not strictly dominated.  Corollary \ref{banningNoMatter2} clarifies that BTB can have an effect only if the testing structure is sufficiently precise.  

Taken together, the two results indicate that ``coarsening'' the employer's information by obscuring an applicant's group membership can have an impact on outcomes only if the employer's information about applicants --- encompassing both his or her prior beliefs about the groups' potentials and his or her interim information about the applicant in question's true qualification --- is sufficiently rich.  Put another way, Corollary \ref{banningNoMatter1} states that the box can have an impact only if the employer might treat workers from the two groups differently even if they are using the same strategy to obtain qualification, and Corollary \ref{banningNoMatter2} states that the box can have impact only if the employer's information about any given worker's qualification is sufficiently precise for the employer to actually condition upon the test result when making his or her hiring decision.  

We now turn to situations in which BTB has an impact on equilibrium outcomes, focusing first on those in which BTB affects \textit{only} $E$'s equilibrium hiring strategy, $\eta$.

\subsection{Situations In Which the Box Affects Only Employer Behavior  \label{Sec:BoxAffectsOnlyEmployerBehavior}} 

By Corollary \ref{banningNoMatter1} we know that, for BTB to have an impact, it must be the case that the groups have statistically distinct potentials ($p_1\geq p_E^* >p_2$).  Figure \ref{Fig:ParetoOptimalEquilibria} illustrates that when $\phi_1\geq \frac{c_L}{w}$ and $\phi_0\geq \frac{c_L}{w}$, an FQE exists when the box is present. Consequently, regardless of whether $E$ hires aggressively or conservatively, all low-cost workers are incentivized to obtain qualification.  However, BTB may affect $E$'s equilibrium hiring strategy.  Because of this, the welfare effects of BTB are ambiguous, as summarized in the following corollary.

\begin{corollary}
\label{Cor:BTBUniformlyInformative}
When the groups are statistically distinct ($p_1\geq p_E^* >p_2$) and the test is uniformly informative ($\min[\phi_0,\phi_1]\geq \frac{c_L}{w}$), employer behavior is affected by the box but worker behavior is not, and BTB has ambiguous welfare effects:
\begin{itemize}
\item If population potential is high ($\overline{p}\geq p_E^*$), BTB induces $E$ to hire all workers aggressively, leaving the payoffs of group 1 workers unchanged and strictly benefiting group 2 workers,
\item If population potential is low ($\overline{p}< p_E^*$), BTB induces $E$ to hire all workers conservatively, strictly hurting group 1 workers and leaving the payoffs of group 2 workers unchanged,
\item Regardless of population potential, $E$ is made strictly worse off by BTB.
\end{itemize}
\end{corollary}

In addition to illustrating that the workers and employer might have opposed preferences about the presence of the box, Corollary \ref{Cor:BTBUniformlyInformative} illustrates the central role of statistical discrimination in our theory by highlighting the importance of the \textit{population potential} for the workers' induced preferences about the box's presence.  When the population at large has high potential, then BTB helps workers in the disadvantaged group and, conversely, when the population has low potential, BTB hurts workers in the advantaged group.  


\paragraph{When the Box Affects Worker and Employer Behavior.} When the box does not affect worker incentives, the only effect the box can have is on employer behavior.  In this case, the employer is always hiring sub-optimally relative to the full-information environment in which $E$ can observe group type, and BTB always makes $E$ worse off.  BTB may make $W$ better or worse off depending on the base rate of potential in the total population relative to $W$'s group potential.

However, when BTB affects worker incentives to obtain qualification, the welfare effects of the box are more nuanced.  In these cases $E$ may strictly prefer to ban the box if doing so can stimulate a greater number of workers to become qualified in equilibrium.  In these situations where $E$ prefers to ban the box it may also be the case that $W$ prefers to ban the box too, and BTB can represent a Pareto improvement.  It can also be the case that BTB stimulates worker qualification and benefits $E$ while hurting $W$.  And finally, it can be the case that BTB can reduce worker incentives to become qualified, leading to losses by both $E$ and $W$.  This final case represents a situation in which observing group labels in the hiring decision is Pareto superior to BTB.

\subsection{BTB With a Positively Informative Test: $\phi_1\geq \frac{c_L}{w}>\phi_0$  \label{Sec:BoxEffectsWithPositivelyInformativeTest}} 

We first consider positively informative test structures.  In such testing structures, the test result is more precise for workers who are qualified ($q=1$) than for workers who are unqualified ($q=0$).  In the Pareto efficient equilibrium in this case, workers in the disadvantaged group ($g=2$) obtain full qualification ($\chi^*=1$) and $E$ hires conservatively from this group ($\eta^*(2)=0$).  On the other hand, the employer $E$ and workers in the advantaged group ($g=1$) are playing mixed strategies in the Pareto efficient equilibrium, as characterized by Equations \ref{Eq:mixedStrategiesEta} and \ref{Eq:mixedStrategiesChi} (substituting the term $p_1$ for $p$ in those equations). $E$ would like to hire aggressively from group 1, but doing so would eliminate $W$'s incentive to obtain qualification because the low $\phi_0$ means that it is likely an unqualified person will send a signal of $2$.  

In such cases, BTB has two potential effects, depending on whether the potential of the population at large, $\overline{p}$, is high or low.  Specifically, when this potential is high, BTB will induce all workers and $E$ to play an MSE, again characterized by Equations \ref{Eq:mixedStrategiesEta} and \ref{Eq:mixedStrategiesChi}, (in this case, substituting the term $\overline{p}$ for $p$ in those equations).  When $\overline{p}< p_E^*$ then BTB will shift all workers to an FQE in which $E$ hires conservatively.  The welfare effects of the box in these cases are not immediately obvious, so we begin with the following lemma.


\begin{lemma}
\label{EUworkersMSE}
Regardless of which group a worker belongs to, and whether the box is used or not, $W$'s expected payoff from the potential mixed strategy equilibrium profile, $(\chi^*,\eta^*)=(\chi_M(p),\eta_M)$, is independent of the worker's realized cost of qualification, $c \in \{c_L,c_H\}$, and equal to the following:
\[
EU_W(\text{MSE}) \equiv
\frac{(1-\phi_0)(\phi_1 w-c_L)}{\phi_1-\phi_0},
\]
while $W$'s conditional expected payoff in a full qualification equilibrium with conservative hiring, given $c \in \{c_L,c_H\}$, is:
\[
EU_W(\text{FQE}\mid \eta^*=0, c) \equiv
\begin{cases} 
\phi_1 w-c_L	& \text{ if }c=c_L\\
0							& \text{ if }c=c_H.
\end{cases}
\]
\end{lemma}

The next proposition establishes that the worker's and employer's induced preferences regarding BTB are opposed when the test is positively informative and population potential is low.
\begin{proposition}
\label{mixedOpposed}
If the test is positively informative ($\phi_1\geq \frac{c_L}{w}>\phi_0$), the groups are statistically distinct ($p_1\geq p_E^*>p_2$), and population potential is low ($\overline{p}<p_E^*$), then $W$ and $E$ have opposed preferences over the box: $E$ prefers that the box be present, $W$ prefers that the box be banned. 
\end{proposition}

On the other hand, in contrast with Proposition \ref{mixedOpposed}, \textit{the worker's and employer's induced preferences regarding BTB are aligned in favor of BTB when the test is positively informative and population potential is high.}  Using the phrase \textit{BTB Pareto dominates the Box} to describe any situation in which there is a Pareto efficient equilibrium without the box that offers both players strictly higher (expected) payoffs than any equilibrium when the box is present, this is stated formally in the following proposition.\footnote{By alluding to \textit{expected} payoffs, we are referring to the worker's expected payoff prior to learning which group he or she is a member of (and, of course, prior to knowing the test result, $\theta$).}
\begin{proposition}  
\label{Pr:BTBParetoEfficientMSE}
If the test is positively informative ($\phi_1\geq \frac{c_L}{w}>\phi_0$), the groups are statistically distinct ($p_1\geq p_E^*>p_2$), and population potential is high ($\overline{p}>p_E^*$), then ``BTB Pareto dominates the Box,'' strictly benefiting $E$ and group 2 workers, and leaving the payoffs of group 1 workers unchanged.
\end{proposition}
Proposition \ref{Pr:BTBParetoEfficientMSE} is one of the key results of our analysis, but we defer discussion of it until Section \ref{Sec:BTBParetoImplications}. 



\subsection{BTB With a Negatively Informative Test: $\phi_0>\frac{c_L}{w}> \phi_1$ \label{Sec:BoxEffectsWithNegativelyInformativeTest}} 

We now consider negatively informative test structures, in which the test result is more precise for unqualified workers than it is for qualified workers.  In the Pareto efficient equilibrium with the box, workers in the disadvantaged group ($g=2$) obtain no qualification ($q=0$). $E$ would hire conservatively from this group ($\eta^*(2)=0$), and consequently the return to investment on qualification is too low to make qualification profitable.  No one in the disadvantaged group becomes qualified, and no one is hired.  Workers in the advantaged group $(g=1)$ obtain full qualification, and $E$ hires from this group aggressively $(\eta^*(1)=1)$. With the box, the payoff for all workers in group 1 is strictly positive: 
\begin{eqnarray*}
w-c_L>0 & \text{ if } & c=c_L \text{ and }\\
w(1-\phi_0)>0 & \text{ if } & c=c_H.
\end{eqnarray*}
The expected payoff for $E$ in this case is 
\[
\gamma(p_1(B-w)-w(1-p_1)(1-\phi_0)) > 0.
\]
This payoff is strictly positive in this case because $p_1\geq p_E^*$ (otherwise BTB has no effect on equilibrium behavior): $E$ receives a strictly positive payoff from hiring individuals from group 1 receiving a test score of $\theta=3$ and a non-negative payoff for hiring individuals receiving a $\theta=2$.

Our first result in this case is that BTB hurts both workers and the employer when population potential is low.
\begin{proposition} 
\label{banningBad}
When the test is negatively informative ($\phi_0>\frac{c_L}{w}> \phi_1$), the groups are statistically distinct ($p_1\geq p_E^*>p_2$), and population potential is low ($\overline{p}<p_E^*$), BTB is Pareto inefficient.
\end{proposition}
Proposition \ref{banningBad} is informative: BTB will reduce employment in equilibrium when the population has low potential.  This can occur for one or more of three reasons: (1) the advantaged workers have moderately high potential ($p_1 \approx p_E^*$), (2) the disadvantaged workers have sufficiently low potential ($p_2$ is too close to zero), and/or (3) the advantaged group is not particularly large ($\gamma$ is too low).  While of course group potentials might vary across different types of jobs, we believe that the third category is the most interesting.  This is because $\gamma$ reflects the proportion of \textit{applicants} for the position in question who come from the advantaged group.  It is well documented that gender, racial, and ethnic compositions of the workforce vary---sometimes widely---across different types of jobs.  Unfortunately, with this in mind, Proposition \ref{banningBad} suggests that BTB may not be as effective at promoting increased employment in sectors that are already disproportionately applied for by citizens from relatively disadvantaged groups.\footnote{Of course, there are many reasons for demographic variation across different jobs, including variation in wages.  Our point here is meant only to be suggestive regarding the empirical implications of our analysis.}

Our second, complementary, result in this case is that BTB benefits workers when population potential is high.

\begin{proposition} 
\label{2likes}
When the test is negatively informative ($\phi_0>\frac{c_L}{w}> \phi_1$), the groups are statistically distinct ($p_1\geq p_E^*>p_2$), and population potential is high ($\overline{p}\geq p_E^*$), BTB strictly benefits group 2 workers and leaves the payoffs of group 1 workers unchanged.
\end{proposition}

Finally, when $\overline{p}\geq p_E^*$ the effect of BTB is ambiguous for the employer, and depends  on whether $E$ receives a positive or negative expected payoff from hiring individuals from group 2 aggressively.  We have assumed that $p_2<p_E^*$, and so it is not sequentially rational for $E$ to hire an individual from group 2 receiving $\theta=2$.  This leads to no qualification by group 2 when $E$ can observe group identity.  However, $E$ may strictly benefit from committing to aggressively hire from this group, because doing so stimulates a full qualification equilibrium.  BTB can serve as a commitment device for $E$ to hire aggressively, when such commitment would not be possible if group identity were observed.  The following proposition details when this commitment benefits $E$.

\begin{proposition} 
\label{paretoE}
When the test is negatively informative ($\phi_0>\frac{c_L}{w}> \phi_1$), the groups are statistically distinct ($p_1\geq p_E^*>p_2$), and population potential is high ($\overline{p}\geq p_E^*$), BTB Pareto dominates the Box if
\[
p_2\in \left[ \frac{w(1-\phi_0)}{B-w\phi_0}, p_E^*\right),
\]
and $E$ is hurt by BTB if 
\[
p_2< \frac{w(1-\phi_0)}{B-w\phi_0}.
\]
\end{proposition}
Again, we defer discussion of this result (along with its sibling, Proposition \ref{Pr:BTBParetoEfficientMSE}) to Section \ref{Sec:BTBParetoImplications}.  Prior to that, we briefly summarize and illustrate the equilibrium effects of BTB. 

\paragraph{The Equilibrium Effects of (and Induced Preferences for) BTB.}  Figure \ref{Fig:ParetoOptimalEquilibria} depicts the equilibrium regions for the cases of low and high group potential (\textit{i.e.}, whether $p$ is less than or greater than $\overline{p}$, respectively).  

\begin{figure}[hbtp]
\centering
\framebox{
\epsfig{file=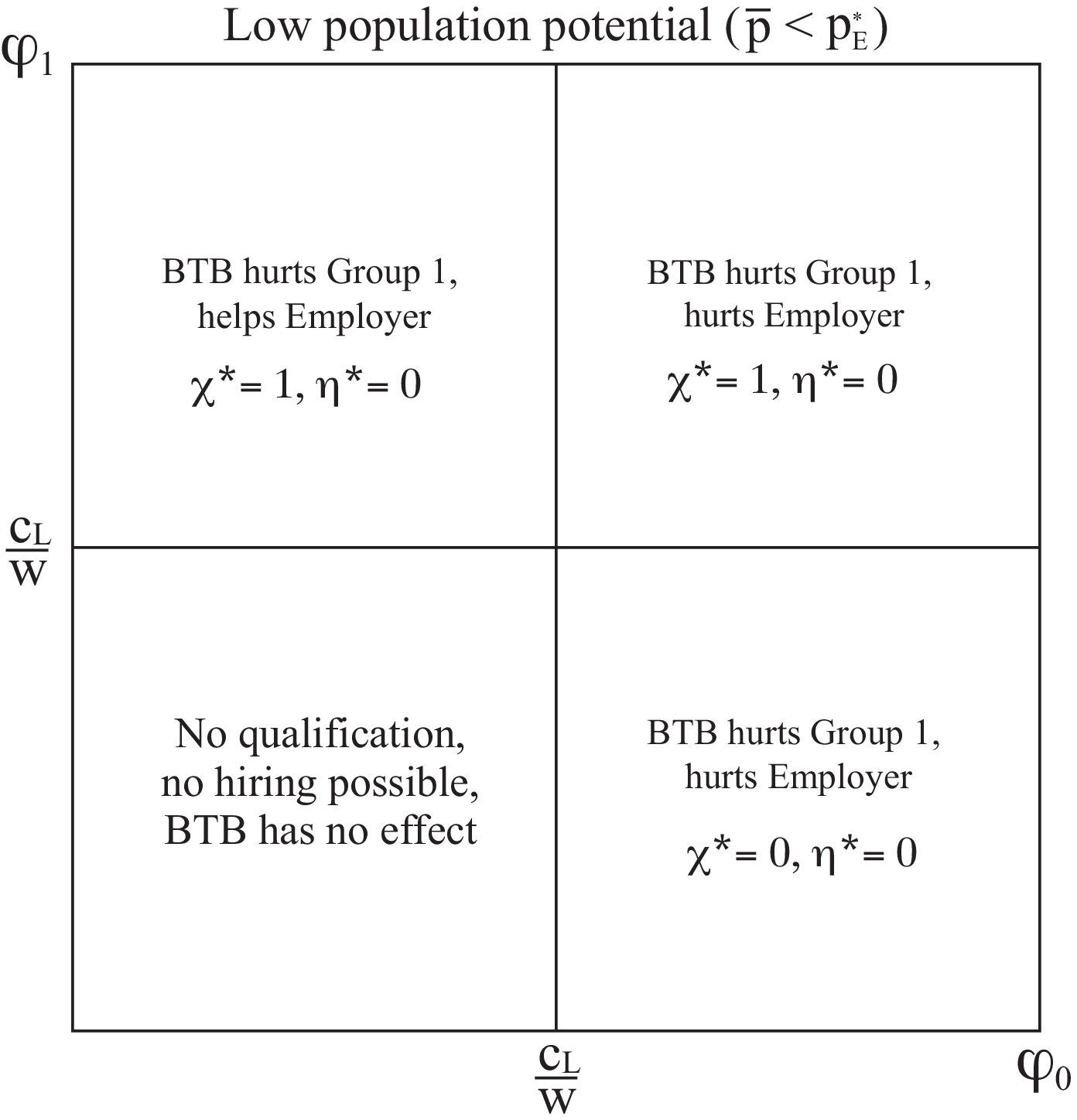, scale=.45} \epsfig{file=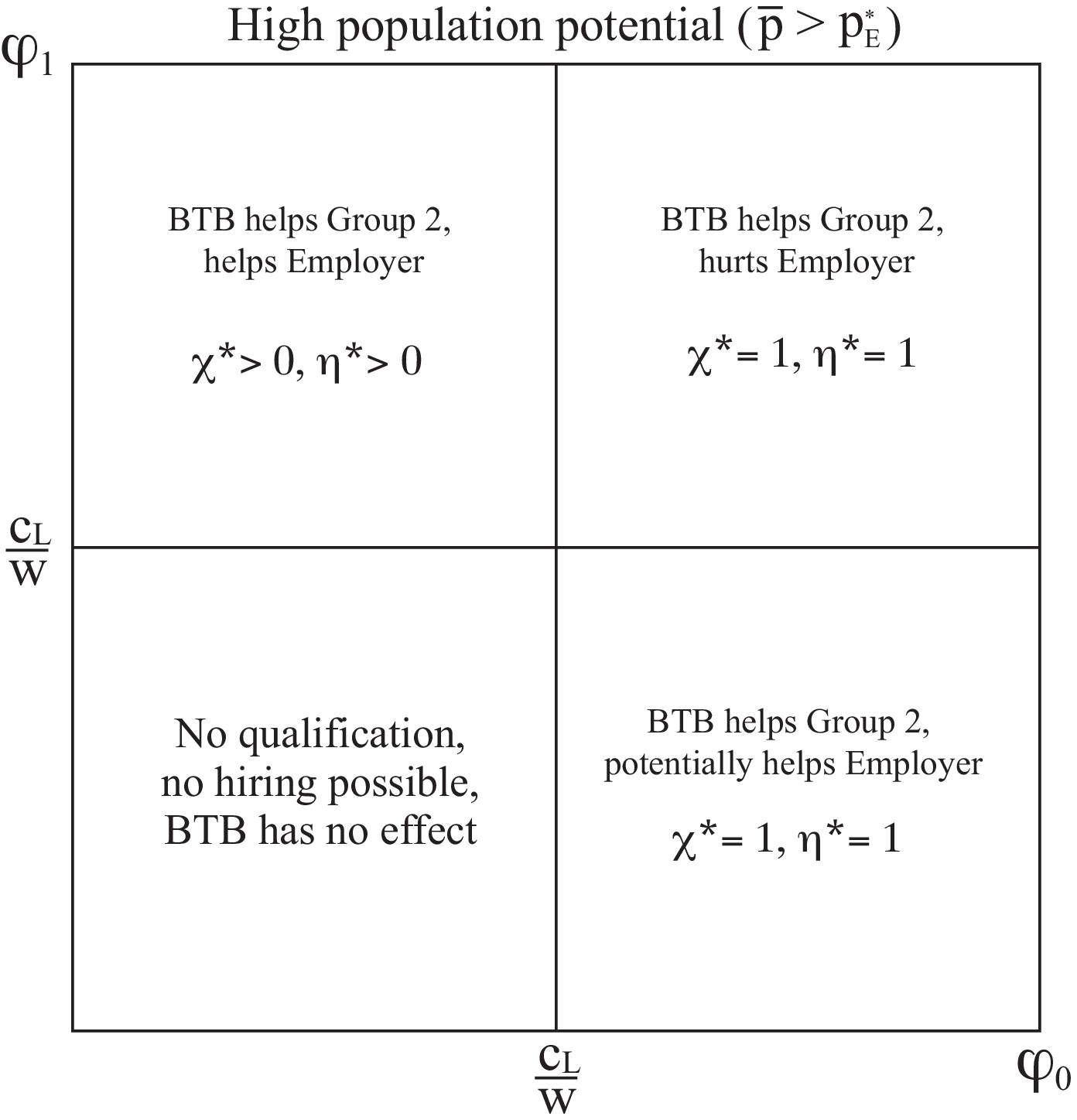, scale=.45}

}
\caption{Equilibrium Effects of BTB When Groups Are Statistically Distinct\label{Fig:EffectOfBTB}}
\end{figure}
Mirroring Figure \ref{Fig:ParetoOptimalEquilibria}, Figure \ref{Fig:EffectOfBTB} depicts regions on which BTB can affect outcomes.  Figure \ref{Fig:EffectOfBTB}  illustrates that the effect of BTB depends critically on whether the population potential, $\overline{p}=\gamma p_1+(1-\gamma)p_2$, is high or low.


\section{BTB and Social Welfare \label{Sec:BTBParetoImplications}}

The finding that BTB can by Pareto dominant is arguably the most provocative of the conclusions we obtain from this framework.  Along these lines, it is informative to contrast Propositions \ref{Pr:BTBParetoEfficientMSE} and \ref{paretoE}.   

\paragraph{Conditional Effects of the Testing Structure.}  Propositions \ref{Pr:BTBParetoEfficientMSE} and \ref{paretoE} are distinguished by the exogenous nature of the testing technology, a point to which we return below (Section \ref{Sec:Empirics}).  They are unified, however, by their common reliance on population potential.  When population potential ($\overline{p}$) is low, BTB can only hurt the workers, but the impact of BTB on the employer is conditional on the testing structure:
    \begin{itemize}
        \item When the test is negatively informative, the employer is hurt by BTB, but
        \item When the test is positively informative, the employer is \textit{helped} by BTB.
    \end{itemize}

This distinction raises the question of what these testing structures represent in substantive terms.  In terms of robustness, it is important to note that our analysis does not rely on the assumption that a qualified worker ($q=1$) can never receive a score of $\theta=1$ or that an unqualified worker ($q=0$) can never receive a score of $\theta=3$: these probabilities can be positive, so long as they are not too large.  With this in mind, one description of the distinction in employment situations is with respect to whether qualifications for the job in question are possible to directly demonstrate (either in one's record or during the hiring process).  
\begin{itemize}
    \item {\bf Negatively Informative Tests.} For many entry-level positions, there are few objective indicators that an individual \textit{is} specifically qualified for the position.  On the other hand, there may be several indicators that an individual is \textit{not} qualified for such a position.\footnote{This asymmetry is due, in part, to the reality that an entry-level position typically does not require that one have held a similar job in the past.  In the modern economy, many such jobs are in retail and customer service positions that do not depend heavily upon task-specific expertise.}    For example, while some ``general'' credentials, such as a high school or college diploma, are relatively easy to verify, they arguably reflect more in their absence than in their presence.  In our model, then, the absence of such a qualification represents $\theta=1$, but a test result of $\theta=3$ would be relatively rare, requiring additional evidence (\textit{e.g.}, a credible and personalized recommendation from a teacher with personal knowledge of the applicant's abilities) that might, but need not, emerge from successfully completing the course of study leading to the diploma.  
    \item {\bf Positively Informative Tests.} As opposed to entry-level positions, more advanced positions often require task-specific experience and skills that can be more easily ``directly'' verified.\footnote{For example, it is arguably easier to reliably infer that an applicant has knowledge of a specific programming language than that the applicant is generally unflappable in a wide array of customer service settings.}  Similarly, for more advanced positions that require specific experience, it is reasonable to suppose that such performance might be gradated into more refined categories, ranging from ``above the bar'' ($\theta=2$) to ``clearly qualified'' ($\theta=3$).  Our analysis requires only that it is rare for a person with the appropriate skills to be identified as ``clearly unqualified.''
\end{itemize}

\paragraph{The Positive Role of BTB for the Employer.}  When the employer benefits from BTB in our framework, it is because BTB is solving a {\bf commitment problem} for the employer.  For example, BTB is playing this role whenever BTB benefits the employer but not the worker (\textit{e.g.}, when population potential is low and the test is positively informative).  This reflects a key simplifying assumption of the model: the worker does not care about qualification, \textit{per se}, and we return to this general point below (Section \ref{Sec:Empirics}) regarding $E$'s incentives to adopt BTB unilaterally. 

\paragraph{Pareto Efficiency \& Inequality.}  Even when BTB is Pareto dominant in our framework, it nonetheless has uneven impacts on welfare.  Comparisons between $W$ and $E$ are difficult for typical reasons (\textit{e.g.}, such a comparison depends on the exogenous parameters $B$ and $w$), but a similar comparison between the two groups of workers is more straightforward.  When BTB is Pareto dominant, its adoption strictly benefits only \textit{disadvantaged} workers.  This mirrors much of the policy and scholarly discussion regarding BTB.  

However, this one-sided nature of the welfare impact of BTB is essentially ``built into'' the model because of our assumption that the two groups of workers differ only in terms of potential (as opposed to, for example, the precision of the testing structure).  Furthermore, the definition of potential is with respect to the employer's payoffs,\footnote{Again, as mentioned above, the worker is assumed to be indifferent with respect to his or her qualification beyond the cost of becoming qualified.} so that any given individual might have a high potential for some employers and low potential for others.  




Some takeaway conclusions from Figure \ref{Fig:EffectOfBTB} can be summarized as follows.
\begin{enumerate}
    \item \textit{BTB has no effect when the test is uninformative} (i.e., \textit{when} $\max[\phi_0,\phi_1]<\frac{c_L}{w}$).  BTB has no effect on outcomes or welfare when the employer's information is too imprecise for the employer to be responsive in equilibrium to the test result, $\theta$. This conclusion is particularly important for ``low information'' situations such as when workers first enter the workforce.  
    
    In our model, the testing structure is exogenous, but in a larger model in which the distribution of $\theta$ is more precise for workers with (say) longer employment histories, the lower left region of Figure \ref{Fig:EffectOfBTB} indicates that BTB policies might not have strong effects on outcomes in labor markets where many of the applicants do not have extensive work histories.  Given that people convicted of a felony in the United States tend to have their first felony conviction before they are 25 years old,\footnote{For example, see Table 7 of \cite{BensonKerley00}.} our analysis offers a reason to temper expectations about the efficacy of BTB in terms of helping convicted felons enter the work force.
    \item \textit{BTB hurts the employer when the test is uniformly informative} (i.e., \textit{when} $\min[\phi_0,\phi_1]>\frac{c_L}{w}$).  In these cases, the worker's moral hazard problem is relatively insignificant from the employer's perspective.  For example, for any $(\phi_0,\phi_1) \approx(1,1)$,  the employer essentially faces \textit{no} moral hazard problem.  In this case, hiding the worker's group membership from the employer can only hurt the employer when groups have statistically distinct potentials (\textit{i.e.}, $p_1>p_E^*>p_2$).
\end{enumerate}

\section{Empirical Implications \label{Sec:Empirics}}

From an empirical standpoint, Propositions \ref{mixedOpposed}--\ref{paretoE} indicate situations in which employers and/or workers would support or oppose BTB.\footnote{For simplicity, in the brief discussion below, we set aside cases where $\phi_0$ and $\phi_1$ are both too low to sustain positive employment in equilibrium. These are identified by the lower left regions in both panes of Figure \ref{Fig:EffectOfBTB}.}  We discuss these incentives briefly, beginning with the two groups of workers and then moving to the employer.  

\paragraph{Workers.}  Regarding the worker's incentives about BTB, the key parameters of interest are (1) whether the worker is advantaged or not and (2) whether population potential is high or low:
\begin{itemize}
    \item Advantaged workers never benefit from BTB and
    \item \textit{When population potential is high}, disadvantaged workers are never harmed by BTB.  
\end{itemize}
From an empirical perspective, it can be useful to think of the two groups of workers (advantaged and disadvantaged) as being distinguished by the likelihood of being hired in the presence of the box.  From this perspective, the conclusion is intuitive: disadvantaged workers have more to gain from BTB than do advantaged workers.  It is important to note that our model considers only two groups.  If we extended the analysis to three or more groups (differentiated, as here, only by their potentials), the ``clean'' conclusions regarding advantaged and disadvantaged workers would apply only to the groups with the highest potential (advantaged) or lowest potential (disadvantaged).  All groups with intermediate potentials could either gain or lose from adoption of BTB.

Of course, workers may have other concerns related to BTB beyond the instrumental motivations focused on here (\textit{e.g.}, altruism, ``expressive'' motivations for BTB, concerns about inequality, \textit{etc}.), so these conclusions should be seen as simply implying that workers from relatively advantaged groups may be less likely to lobby for adoption of BTB than workers from disadvantaged groups.  At the industry or occupation levels, this implies that one might see more active lobbying for BTB policies in jobs or sectors that tend to have higher proportions of applicants who come from disadvantaged groups.  This logic implies that jobs that require significant work experience (or, perhaps verified educational attainment) will tend to receive less focus in policy debates regarding BTB.  At least on a cursory glance, this jibes with real-world experience: BTB policies appear to be most actively lobbied for in entry-level and/or early career positions.

Similarly, in terms of the workers' potentially divergent interests regarding BTB, it is important to note that our model's minimal structure implies that it is never the case that one group of workers will strictly benefit from, while the other group is strictly harmed by, BTB.  If we interpret ``is not harmed by BTB'' as ``might be harmed by BTB'' and ``is not helped by BTB'' as ``might benefit otherwise from BTB,'' then the model readily identifies situations in which there may be conflict about BTB among workers.  Indeed, the theory indicates that such conflict might be quite common --- in any situation in which there is an equilibrium with the box banned in which workers get hired with a positive probability, the advantaged workers can only be harmed by BTB and the disadvantaged workers can only benefit from it.  Such a possibility seems even more likely to emerge if we allowed the employer to choose from two applicants, with randomly determined group memberships.  \label{Pg:CompetitiveHiring} In such a ``competitive'' hiring model, the advantaged workers will strictly benefit from the box in expectation and the disadvantaged workers will be similarly, strictly, harmed by the box's presence.  Even beginning the analysis of such an extension is clearly beyond the scope of this article, but the possibilities apparent in such an extension seem worthy of exploration.

\paragraph{Employers.}  One of the most surprising aspects of our analysis is its conclusion that BTB can sometimes benefit the employer. 
One substantive implication from the analysis is that the employer always prefers BTB when the test is positively informative, but not necessarily when the test is negatively informative.

Though BTB's effects depend on population potential, the employer's preferences for BTB are independent of population potential when the test is positively informative.  From this, one might expect that employers will be more receptive of BTB policies --- perhaps even committing to BTB unilaterally --- for positions in which qualification is relatively likely to be clearly indicated through the test result (\textit{i.e.}, $\theta=3$).  This might induce conflict between the employer and workers --- particularly workers from the advantaged group --- when population potential is low.\footnote{An interesting ancillary implication of this --- related to the question of voluntary disclosure (discussed in Section \ref{Sec:Conclusion} on page \pageref{Pg:VoluntaryDisclosure}, below) --- is that workers who seek to ``signal their group membership'' through their (unmodeled here) lobbying efforts for/against BTB may have an incentive to lobby insincerely: disadvantaged workers might lobby \textit{against} BTB if they believe that (1) BTB might actually be adopted and (2) the employer might observe this lobbying effort and thereby possibly incorporate this lobbying effort into $E$'s beliefs about $W$'s group membership if $W$ ever applies for a job from $E$.  Of course, such behavior is consistent with other dynamic motivations, but the congruence is nonetheless intriguing.}


When the test is negatively informative, on the other hand, the employer can gain from BTB only if population potential is high.  From an empirical perspective, this implies that employers should be more receptive to BTB policies for positions in which relatively few applicants are from the disadvantaged group (otherwise, population potential would be low if the groups have statistically distinct potentials).  Accordingly, the employer's interests in BTB are not entirely in line with social welfare maximization.

This partial divergence between $E$'s incentives and social welfare is further exacerbated by the fact that $E$'s preference for BTB when the test is negatively informative is conditional: from Proposition\ref{paretoE}, $E$ will prefer BTB in this case only if the disadvantaged group's potential is not too low:
\begin{equation}
\label{Eq:EmployerImplications}
p_2\in \left[ \frac{w(1-\phi_0)}{B-w\phi_0}, \frac{w(1-\phi_0)}{B(1-\phi_1)+w(\phi_1-\phi_0)}\right),
\end{equation}
where we have replaced the upper bound, $p_E^*$, with its formal definition.

From \eqref{Eq:EmployerImplications}, it follows that --- for jobs in which the test structure is negatively informative --- employers should be more likely to support/adopt BTB when 
\begin{enumerate}
    \item the market wage, $w$, is low,
    \item the employer's gain from successful hiring ($B$) is large, and/or
    \item the test is very likely to clearly reveal \textit{lack} of qualification (high $\phi_0$).
\end{enumerate}

These conclusions are partially congruent with the fact that much of the discussion of BTB ``as policy'' (\textit{e.g.}, where some wish to impose BTB on private employers, as opposed to employers voluntarily adopting BTB) revolves around entry level positions, where indicators of qualification are frequently imprecise.  Formally, if $(\phi_0,\phi_1) \to (0,0)$ (a completely uninformative testing structure), the lower and upper bounds of \eqref{Eq:EmployerImplications} converge to $\frac{w}{B}$.  Eventually, any such sequence of probabilities will support only zero qualification equilibria (the lower left corner region of the plots in Figure \ref{Fig:EffectOfBTB}), but regardless, the employer will not support BTB in the limit.  Note that, in spite of this, disadvantaged workers will never oppose BTB in the sequence of situations determined by $\{(\phi_0,\phi_1)_t\}_{t=1}^\infty$.  Thus, the theory suggests that a potential cause for conflict about BTB is severe informational imperfection in the moral hazard problem facing $E$ and $W$ in the hiring process.

\section{Discussion, Extensions, and Conclusions \label{Sec:Conclusion}}

When discrimination occurs due to disparate treatment, the decision-maker must be able to observe or infer others' group memberships.  We have examined how and whether eliminating this information, which necessarily eliminates one form of discrimination, might affect qualification, employment rates, and welfare.  This approach is distinct from traditional approaches to eliminating discrimination, which take as given that the employer has access to the information required for discrimination.  

Policies intended to eliminate discrimination on the basis of group membership typically prohibit hiring procedures that explicitly utilize group information; these policies prohibit disparate treatment.  A response taken by some to abide by such prohibitions is to not collect information on group membership.\footnote{Clearly, for any given employment decision, certain group memberships are appropriate considerations for the employer (for example, does the applicant have a high school diploma?).  Accordingly, in practice, discrimination is legally barred only with respect to certain group memberships.  For example, in the United States, federal anti-discrimination laws generally protect against discrimination on the basis of race, color, national origin, religion, sex, age, or disability.}  However, in some cases, collection of this information is unavoidable, either by direct observation or by inference from other information about the applicant, such as their date of graduation, the schools they have attended, and so forth.

Less closely related are requirements that the decisions not be (too strongly) correlated with applicants' group memberships (\textit{i.e.}, prohibiting disparate impact).  Such prohibitions are arguably more appropriately aimed at outcomes, rather than process.  However, they can have spillover effects, whereby attempts to protect one group may reduce welfare for \textit{all} individuals (\textit{e.g.}, \cite{CoateLoury93}).

\paragraph{Discrimination and Big Data.}  While we have framed the discussion of the model within the context of the Ban the Box movement, it of course has implications for information about any characteristics of the applicants.  Furthermore, the framework can easily be extended to incorporate noisy signals about group membership, so that the employer must form non-degenerate beliefs after observing some exogenous, imperfectly informative information about the applicant's characteristics.  Particularly in the new age of algorithms and ``big data,'' the data solicited for decision-making can have subtle and powerful impacts on outcomes (\cite{PattyPenn15PS,KleinbergLudwigMullainathanSunstein18}).  The power of certain information is of course not new --- it doesn't require a supercomputer for an employer to discriminate against employees based on race, gender, criminal record, or any other single factor.  However, with massive, and often proprietary, data sets and algorithms, it is much more difficult to predict which subset of seemingly innocuous questions might be, either explicitly or implicitly, used as the basis for discrimination (\cite{BarocasSelbst16}).  Such discrimination can emerge in various settings, including college admissions, employment, purchasing insurance, and obtaining credit.\footnote{Beyond the scope of this article, but related, is the emerging topic of how algorithmic systems may produce \textit{disparate mistreatment}: situations in which the algorithm's decisions are more accurate for one group than for another (\cite{ZafarValeraGomezRodriguezGummadi17}).}

\paragraph{Extensions.}  The model has several avenues for extension.  We briefly describe five of these below.

\begin{enumerate}
    \item \textbf{Endogenous Wages.} The analysis in this article assumes that the employer must offer an exogenously determined wage when he or she hires a worker.  We are currently relaxing this assumption in ongoing work, but several constraints preclude us from reporting the preliminary results in detail.  One important fact that can be inferred from the analysis reported in this article is that the employer has less ``need for'' information about an applicant's group membership if the employer can choose (and commit to) a wage prior to the worker choosing whether to get qualified.  This can be seen in several places, but perhaps the most transparent is Figure \ref{Fig:ParetoOptimalEquilibria}, in which the horizontal and/or vertical lines in the two panes are each a function of the wage, $w$.
    \item \textbf{Group-Specific Testing Accuracy.}  Our analysis above assumes that the only distinction between the two groups is their potential, a notion grounded in inherent opportunities available to the individuals in the two groups.  A complementary analysis would consider the implications of the testing technology (\textit{i.e.}, the distribution of $\theta$  conditional on qualification, $q$) depending on the worker's group.  Such an analysis would be interesting for several reasons, including raising the possibility that banning the test itself might be socially optimal.
    \item \textbf{Intersectionality.}  Our analysis focuses on the case in which there are two (observable) groups of workers.  Reality is of course more complicated: there are many forms of group membership that are relevant in the awarding of selective benefits (\textit{e.g.}, race, ethnicity, citizenship, age, gender, and veteran status).  Most interesting about such an extension is the potential to explore the implications and challenges of issues of \textbf{intersectionality} ---``the way in which various forms of inequality often operate together and exacerbate each other''\footnote{Kimberl\'{e} Crenshaw, quoted in ``She Coined the Term `Intersectionality' Over 30 Years Ago. Here's What It Means to Her Today,'' by Katy Steinmetz, \textit{TIME}, February 20, 2020. For a very recent formal contribution along these lines, see \cite{Stewart21}.} --- when considering the impact of supplemental information on allocating scarce resources.
    \item \textbf{Voluntary Disclosure.}  \label{Pg:VoluntaryDisclosure} Our analysis is centered on the effects of information about an individual's traits.  In reality, this information is often solicited by the employer, as opposed to being directly observed.  Accordingly, an important extension of the model would be to include voluntary provision/revelation of this information within the model itself.
    \item \textbf{Competitive Hiring.}  As mentioned in Section \ref{Sec:Empirics} (pg. \pageref{Pg:CompetitiveHiring}), one direction to extend the model is to incorporate the possibility that the employer will have a larger set of applicants to choose from than the number of positions he or she needs to hire.  This extension, as mentioned above, would appear to induce strict preferences for the workers about BTB in a way that is richer than captured in the model analyzed here and could offer a useful springboard for analyzing under what conditions (say) a government might find it in its (political, electoral, and/or economic) interest to impose BTB as a matter of public policy.
\end{enumerate}

\paragraph{Concluding Thoughts.}  We have presented a highly stylized model of hiring with moral hazard with the aim of considering the impact of heterogeneity among workers in terms of the opportunity to become qualified and, more specifically, the impact of the employer's granular information about this heterogeneity in the hiring process.  Policies such as Ban the Box are aimed squarely at ``leveling the playing field'' for individuals from different backgrounds.  Our analysis indicates some of the promises --- and pitfalls --- of such policies.  In line with empirical evidence, the theory highlights the generally positive impact such policies will have on disadvantaged workers and the weakly negative impact they might have on advantaged workers.

In contrast, the theory also indicates unsurprisingly that such policies can sometimes harm employers, while at the same time offering (to us, at least) an unexpected conclusion: sometimes employers can strictly benefit from these policies if they are foreseen and reacted to by workers in the disadvantaged group(s).  Furthermore, the theory isolates one classic game theoretic reason for this potential salutary impact: the employers in some cases benefit from the ``ignorance'' imposed on the employer by such policies because the concomitant lack of ability for the employer to discriminate between workers from the two groups can provide instrumental incentives to workers from such groups to make costly investments in qualification in the hopes of obtaining employment on the now-leveled ``playing field.''

While our model omits many interesting features of real-world employment markets, we think that the minimalism of the model highlights the ubiquity of the potential Pareto efficiency of partially ``blinding'' decision-makers engaged in distributing rewards among citizens.
\newpage

\appendix

\section{Proofs \label{Sec:Proofs}}

\noindent {\bf Proposition \ref{Pr:EquilibriumCharacterization}}  \textit{The following table characterizes all equilibria in which positive qualification can be obtained. When multiple equilibria exist, they are strictly Pareto ranked.}
\[
\begin{array}{|c|c|}
\multicolumn{2}{c}{\text{Equilibria when }p> p_E^*}\\ \hline
\text{Parameters (} c_L,w,\phi_0,\phi_1\text{)} & \text{Equilibria} \\
\hline
w \phi_0>c_L>w\phi_1 & 
\begin{array}{c}
\text{FQE with }\chi^*=1, \eta^*=1,\\
\text{MSE with }\chi^*=\chi_M(p), \eta^*=\eta_M,\\
\text{ZQE with }\chi^*=0, \eta^*=0,
\end{array}
\\ \hline
w\phi_0>c_L\text{ and }w\phi_1>c_L & \text{FQE with }\chi^*=1, \eta^*=1 \\\hline
w \phi_1>c_L>w\phi_0 & \text{MSE with }\chi^*=\chi_M(p), \eta^*=\eta_M,\\ \hline
c_L>w\phi_0\text{ and }c_L>w\phi_1 & \text{ZQE with }\chi^*=0, \eta^*=0 \\ \hline
\multicolumn{2}{c}{}\\
\multicolumn{2}{c}{\text{Equilibria when }p< p_E^*}\\ \hline 
\text{Parameters (} c_L,w,\phi_0,\phi_1\text{)} & \text{Equilibria} \\ 
\hline
c_L>w\phi_1 & \text{ZQE with }\chi^*=0, \eta^*=0  \\ \hline
w\phi_1>c_L & \text{FQE with }\chi^*=1, \eta^*=0 \\\hline
\end{array}
\]
\begin{proof} 
We proceed through the six regions identified in the statement of the proposition.  For the first four cases, note that when $p>p_E^*$, as defined in \eqref{Eq:OverlineP}, $E$ receives a strictly positive payoff from hiring $\theta=2$ if all low types have chosen qualification.  When $p<p_E^*$ then $E$ receives a strictly negative payoff from hiring $\theta=2$ if all low types have chosen qualification, and $E$ will consequently never hire if observing $\theta=2$.\\

\noindent {\sc Region 1.} $p>p_E^*$ and $w \phi_0>c_L>w\phi_1$. Because $p>p_E^*$, $\eta=1$ is a unique best response to $\chi=1$.  As $\phi_0 w>c_L$, Equation \eqref{Eq:ICHire2} is satisfied and $\chi=1$ is a unique best response to $\eta=1$.  Consequently, there is an FQE with $\eta^*=1$ and $\chi^*=1$ and no other pure strategy equilibrium with full qualification.  It is straightforward to verify that when $p>p_E^*$ and $w \phi_0>c_L>w\phi_1$, then $\eta_M\in(0, 1)$ and $\chi_M(p)\in(0, 1)$, where $\eta_M$ and $\chi_M(p)$ are characterized by Equation \eqref{Eq:mixedStrategiesEta} and \eqref{Eq:mixedStrategiesChi}.  Therefore there also exists a mixed strategy equilibrium for this parameter region.  Finally, Equation \ref{Eq:ICHire3} does not hold in this case, as $c_L>w\phi_1$ and $\chi=0$ is a best response to $\eta=0$.  It follows that there also exists a zero qualification equilibrium in this region.\\

\noindent {\sc Region 2.}  $p>p_E^*$ and $w\phi_0>c_L\text{ and }w\phi_1>c_L $.  As in the above case, $\eta^*=1$ is a unique best response to $\chi^*=1$ and vice versa because $p>p_E^*$ and $\phi_0 w>c_L$.  However in this case there is no MSE, because when $w\phi_1>c_L $ and $w \phi_0>c_L$, $E$ can't choose a hiring strategy $\eta$ to make $W$ indifferent between qualification and no qualification.  Regardless of $E$'s hiring strategy, it is always strictly optimal for $W$ to choose $q=1$.\\

\noindent {\sc Region 3.} $p>p_E^*$ and $w \phi_1>c_L>w\phi_0 $.  In this case, there does not exist a pure strategy equilibrium.  If $\chi=1$ then $E$ optimally chooses $\eta=1$, as $p>p_E^*$.  However, Equation \eqref{Eq:ICHire2} does not hold; when $E$ hires those receiving $\theta=2$ ``aggressively" (i.e. $\eta=1$) and when $\phi_0$ is sufficiently low, $W$ is incentivized to not obtain qualification.  However, if $W$ obtains no qualification then $E$ will not hire if observing $\theta=2$.  In this case there is only a mixed strategy equilibrium, and again it is straightforward to verify that when $p>p_E^*$ and $w \phi_1>c_L>w\phi_0 $ then $\eta_M\in(0, 1)$ and $\chi_M(p)\in(0, 1)$.\\

\noindent {\sc Region 4.}   $c_L>w\phi_0\text{ and }c_L>w\phi_1$. In this case Equation \ref{getQ} can never obtain for any value of $\eta$.  It follows that $\chi^*=0$ and $\eta^*=0$ is the unique equilibrium.\\

\noindent {\sc Region 5.} $p<p_E^*$ and $c_L>w\phi_1$. In these remaining two cases $E$ always sets $\eta^*=0$, because $p<p_E^*$.  When $c_L>w\phi_1$ then Equation \eqref{Eq:ICHire3} does not hold, and $W$ sets $\chi^*=0$.\\

\noindent {\sc Region 6.}  $p<p_E^*$ and $w\phi_1>c_L$. In this last case Equation \eqref{Eq:ICHire3} \textit{does} hold, and $W$ sets $\chi^*=1$.\\

\noindent {\sc Pareto Ranking Equilibria in Region 1.}  We conclude by ranking the 3 equilibria in Region 1 ($w \phi_0>c_L>w\phi_1 $ and $p>p_E^*$) according to the Pareto principle.  It is straightforward to show that the full qualification equilibrium with $\chi^*=1, \eta^*=1$ Pareto dominates the mixed strategy equilibrium. To see this, note that at an MSE, the low-cost worker must be indifferent between obtaining qualification and not, and the employer must be indifferent between hiring a worker with $\theta=2$ and not, and so must receive an expected payoff of zero conditional on $\theta=2$.  However, when $p>p_E^*$, $E$ receives a strictly positive payoff in the FQE from hiring a worker with $\theta=2$. Moreover, at the FQE there is a higher probability a randomly drawn worker will receive a $\theta=3$ (as there is a higher probability $q=1$), and a lower probability that a randomly drawn worker will receive $\theta=0$.  Thus, $E$ receives a strictly higher expected payoff in the FQE than in the MSE.

In the MSE,  a low-cost worker receives a (positive) expected payoff of $(1-\phi_o)w\cdot =\frac{(1-\phi_0)(c_L-w \phi_1 )}{\phi_0-\phi_1}$. In the FQE, $W$ receives an expected payoff of $w-c_L$.  The difference between these payoffs is 
\[
w-c_L-\frac{(1-\phi_0)(c_L-w \phi_1 )}{\phi_0-\phi_1}=\frac{(1-\phi_1)(\phi_0 w-c_L )}{\phi_0-\phi_1},
\]
which, by inspection, is strictly positive when $w\phi_0>c_L>w\phi_1$.  Therefore,  a low-cost worker strictly prefers the FQE to the MSE when both equilibria exist and high-cost workers also strictly prefer the FQE to the MSE, because at the FQE the employer is hiring all workers who receive $\theta=2$, which strictly benefits workers who are not qualified. Finally, note that both the FQE and MSE are strictly Pareto superior to the ZQE, in which both players receive a payoff of 0 with certainty. 

\end{proof}

\noindent {\bf Lemma \ref{EUworkersMSE}} \textit{Regardless of which group a worker belongs to, and whether the box is used or not, $W$'s expected payoff from the potential mixed strategy equilibrium profile, $(\chi^*,\eta^*)=(\chi_M(p),\eta_M)$, is independent of the worker's realized cost of qualification, $c \in \{c_L,c_H\}$, and equal to the following:}
\[
EU_W(\text{MSE}) \equiv
\frac{(1-\phi_0)(\phi_1 w-c_L)}{\phi_1-\phi_0},
\]
\textit{while $W$'s conditional expected payoff in a full qualification equilibrium with conservative hiring, given $c \in \{c_L,c_H\}$, is:}
\[
EU_W(\text{FQE}\mid \eta^*=0, c) \equiv
\begin{cases} 
\phi_1 w-c_L	& \text{ if }c=c_L\\
0							& \text{ if }c=c_H.
\end{cases}
\]
\begin{proof} Any mixed strategy equilibrium is characterized by Equations \ref{Eq:mixedStrategiesEta} and \ref{Eq:mixedStrategiesChi}, with group potential $p$ varying depending on the group being considered (i.e. whether it is a subgroup with potential $p_g$ or the set of all workers with potential $\overline{p}$).  Note that $E$'s mixed strategy, $\eta^*_M$, is not a function of group potential.  $E$ is simply making $W$ indifferent between qualification and no qualification, and this indifference is solely dependent on costs to qualification, wages, and the testing technology, all of which are invariant to the presence or absence of the box.\footnote{Note that, conditional on $W$ and $E$ playing the MSE, $W$ is indifferent about his or her cost of becoming qualified, $c$.  This is because, in our setting, a worker with low costs of qualification is essentially choosing whether to ``act like he or she must have a low cost of qualification'' ($q=1$) or ``act like he or she might have had a high cost of qualification'' ($q=0$).  The worker has a strict preference in equilibrium for a low cost of qualification only in an FQE.}

Any worker playing an MSE will receive an expected payoff of 
\[
\chi_M(w(\phi_1+(1-\phi_1)\eta_M)-c)+w(1-\chi_M)(1-\phi_0)\eta_M,
\]
which reduces to $\frac{(1-\phi_0)(\phi_1 w-c)}{\phi_1-\phi_0}$ (\textit{i.e.}, the group's potential, $p$, drops out of the equation).  Therefore the MSE payoff to the worker is independent of the worker's group identity or the presence or absence of the box.

Finally, in any FQE with conservative hiring an unqualified worker receives a payoff of zero and a qualified worker receives a payoff of $\phi_1 w-c_L$.  In this case expected payoffs again are independent of group identity. 
\end{proof}

\noindent {\bf Proposition \ref{mixedOpposed}}
\textit{
If the test is positively informative ($\phi_1\geq \frac{c_L}{w}>\phi_0$), the groups are statistically distinct ($p_1\geq p_E^*>p_2$), and population potential is low ($\overline{p}<p_E^*$), then $W$ and $E$ have opposed preferences over the box: $E$ prefers that the box be present, $W$ prefers that the box be banned. 
}
\begin{proof}
When $\overline{p}<p_E^*$ then banning the box will generate an FQE with conservative hiring of all individuals. By Lemma \ref{EUworkersMSE} we know that $W$ always prefers the MSE to the FQE with conservative hiring. By the supposition that $\phi_1 w-c_L>0$ and $\phi_0<\phi_1\leq 1$, this follows from the fact that 
\[
\frac{(1-\phi_0)(\phi_1 w-c_L)}{\phi_1-\phi_0} \geq \phi_1 w-c_L > 0.  
\]
Accordingly, both high and low-cost workers prefer the MSE, implying that workers in group 1 are made strictly worse off with the box, and workers in group 2 are indifferent about the box's presence.

The employer's payoff is affected by BTB solely through the change induced in group 1's qualification strategy behavior by BTB, because $E$ was previously at a conservative hiring FQE with group 2 when the box was present.  With the box, $E$ received an expected payoff from hiring from group 1 equal to:
\[
p_1((B-w)\chi_M(\phi_1+(1-\phi_1)\eta_M)-w(1-\chi_M)(1-\phi_0)\eta_M)-w(1-p_1)(1-\phi_0)\eta_M.
\]
This can be reduced to 
\begin{equation}
    EU_E(\text{MSE}|g=1)=\frac{(1-\phi_0)\phi_1(B-w)w}{B(1-\phi_1)+w(\phi_1-\phi_0)}. \label{Eq:EmployerPayoffMSE}
\end{equation}
At the FQE, $E$'s expected payoff from hiring from group 1 is 
\[
EU_E(\text{FQE}, \eta^*=0|g=1)=p_1(\phi_1)(B-w).
\]
Comparing these two payoffs we get that:
\[
EU_E(\text{FQE}, \eta^*=0|g=1)\geq EU_E(\text{MSE}|g=1)
\]
when 
\[
\phi_1(B-w) \left(p_1-\frac{w(1-\phi_0)}{B (1-\phi_1)+w (\phi_1-\phi_0)}\right)\geq 0,
\]
or 
\[
\phi_1(B-w) \left(p_1-p_E^*\right)\geq 0.
\]
Since we have supposed that $p_1>p_E^*$, this inequality always holds.  Therefore $E$ receives a weakly higher payoff (strictly higher if $p_1>p_E^*$) from banning the box in this case. 
\end{proof}

\noindent {\bf Proposition \ref{Pr:BTBParetoEfficientMSE}}
\textit{
If the test is positively informative ($\phi_1\geq \frac{c_L}{w}>\phi_0$), the groups are statistically distinct ($p_1\geq p_E^*>p_2$), and population potential is high ($\overline{p}>p_E^*$), then BTB is Pareto dominant, strictly benefiting $E$ and group 2 workers, and leaving the payoffs of group 1 workers unchanged.
}
\begin{proof}
Note that workers from group 1 are indifferent about banning the box when $\overline{p}>p_E^*$.  First $E$'s mixed equilibrium strategy, $\eta_M$, is unchanged regardless of whether the box is present or not.  Second, with or without the box, workers in the advantaged group play an MSE with $E$.  We demonstrate the result by considering each groups of workers in turn, followed by the employer.

{\sc Workers in the Advantaged Group.} The equilibrium probability that a low-cost worker becomes qualified when the box is banned, $\chi_M(\overline{p})$, is higher than it is for workers from group 1 when the box is present.  This is because the population's potential, $\overline{p}$, is less than $p_1$ and therefore these workers must become qualified at a higher rate in order to keep $E$ indifferent in the absence of the box when considering whether to hire a worker who received a test score of $\theta=2$.  However, this higher rate of qualification by workers from group 1 has no effect on their equilibrium expected payoffs because these workers are indifferent between qualification and no qualification in equilibrium regardless of the box's presence:
\begin{eqnarray*}
EU_W(\text{MSE}\mid g=1) & = & p_1 \cdot \bigg(\chi_M(\overline{p}) (\phi_1 w + (1-\phi_1) \eta_M - c_L) + (1-\chi_M(\overline{p}))(1-\phi_0) \eta_M w\bigg) \\
& & + (1-p_1) (1-\phi_0) \eta_M w,\\
& = & p_1 \cdot \bigg((1-\phi_0) \eta_M w\bigg) + (1-p_1) (1-\phi_0) \eta_M w,\\
& = & (1-\phi_0) \eta_M w.
\end{eqnarray*}

{\sc Workers in the Disadvantaged Group.} Turning to workers in group 2, Lemma \ref{EUworkersMSE} implies that both low-and high-cost workers in group 2 receive a strictly higher payoff in the MSE.  This is the Pareto efficient equilibrium if the box is banned, implying that workers from group 2 strictly benefit from banning the box.

{\sc The Employer.} The employer strictly prefers to ban the box in this setting.  His or her payoff from banning the box in the MSE is equivalent to his or her payoff from the workers that send $\theta=3$; this is because for $E$ to mix conditional on $\theta=2$, $E$ must be receiving an expected payoff of zero conditional on $\theta=2$.  Consequently, $E$'s expected payoff with the box banned is 
\begin{eqnarray*}
EU_E(\text{MSE}) & = & (B-w)\left(\gamma p_1 \chi_M(\overline{p})\phi_1   +(1-\gamma)p_2 \chi_M(\overline{p})\phi_1 \right),\\
& = & (B-w) \cdot \overline{p} \cdot  \chi_M(\overline{p}) \phi_1,\\
& = & (B-w)\phi_1 p_E^*.
\end{eqnarray*}
In the presence of the box, the employer's expected payoff in the Pareto efficient equilibrium in this case is
\begin{eqnarray*}
EU_E(\text{FQE}) & = & (B-w)\left(\gamma p_1 \chi_M(p_1)\phi_1   +(1-\gamma)p_2 \phi_1 \right),\\
& = & (B-w)\left(\gamma p_1 
\frac{w(1-\phi_0)}{p_1(B(1-\phi_1)+w(\phi_1-\phi_0))}
\phi_1   +(1-\gamma)p_2 \phi_1 \right),\\
& = & 
(B-w)\phi_1 \left(\gamma p_E^*
+(1-\gamma)p_2 \right).
\end{eqnarray*}
Accordingly, by the supposition that $p_2<p_E^*$, it follows that $\gamma p_E^* + (1-\gamma)p_2< p_E^*$, implying that $EU_E(\text{MSE})>EU_E(\text{FQE})$, so that $E$ strictly benefits from BTB.

Thus, relative to the expected payoff from the Pareto efficient equilibrium with the box present, the expected payoff from the Pareto efficient equilibrium with BTB is 
\begin{enumerate}
    \item identical for workers from the advantaged group,
    \item strictly higher for workers from the disadvantaged group, and
    \item strictly higher for the employer.
\end{enumerate}
Accordingly, BTB is Pareto dominant in this case, as was to be shown.
\end{proof}

\noindent {\bf Proposition \ref{banningBad}} \textit{When the test is negatively informative ($\phi_0>\frac{c_L}{w}> \phi_1$), the groups are statistically distinct ($p_1\geq p_E^*>p_2$), and population potential is low ($\overline{p}<p_E^*$), BTB is Pareto inefficient.
}
\begin{proof}
When $\overline{p}<p_E^*$, the inequality in Equation \ref{Eq:Hire2OneGroup} fails to hold and $\eta^*(\emptyset)=0$; $E$ hires conservatively from the group at large.  As Inequality \ref{getQ} doesn't hold when $\frac{c_L}{w}> \phi_1$, it follows that $\chi^*(1)=\chi^*(2)=0$ and  the effect of banning the box is to shut the labor market down entirely. No worker obtains qualification, and no worker is hired.  This leaves payoffs for workers in group 2 unchanged.   $E$ and workers in group 1 are strictly worse off than they were with the box.  As described above, with the box both $E$ and workers from group 1 received a strictly positive expected payoff. 
\end{proof}

\noindent {\bf Proposition \ref{2likes}}
\textit{When the test is negatively informative ($\phi_0>\frac{c_L}{w}> \phi_1$), the groups are statistically distinct ($p_1\geq p_E^*>p_2$), and population potential is high ($\overline{p}\geq p_E^*$), BTB strictly benefits group 2 workers and leaves the payoffs of group 1 workers unchanged.}
\begin{proof} 
By satisfaction of Equation \ref{Eq:Hire2OneGroup}, $E$ hires aggressively from the group at large when $\overline{p}\geq p_E^*$.  And by satisfaction of Equation \ref{Eq:ICHire2}, all workers obtain qualification and an FQE exists with $\eta^*(\emptyset)=\chi^*(1)=\chi^*(2)=1$.  The payoff to  members of group 1 at this equilibrium is identical to their payoff when the box was present.  However every member of group 2 is strictly better off in expectation.  With the box, all members of group 2 received a payoff of zero.  Without the box, high cost individuals in group 2 receive an expected payoff of $w(1-\phi_0)>0$ and low-cost individuals receive a payoff of $w-c_L>0$. 
\end{proof}

\noindent {\bf Proposition \ref{paretoE}} 
\textit{When the test is negatively informative ($\phi_0>\frac{c_L}{w}> \phi_1$), the groups are statistically distinct ($p_1\geq p_E^*>p_2$), and population potential is high ($\overline{p}\geq p_E^*$), BTB is Pareto dominant if
\[
p_2\in \left[ \frac{w(1-\phi_0)}{B-w\phi_0}, p_E^*\right),
\]
and $E$ is hurt by BTB if 
\[
p_2< \frac{w(1-\phi_0)}{B-w\phi_0}.
\]
}
\begin{proof}
Banning the box strictly benefits $E$ when $E$'s expected payoff from hiring members of group 2 aggressively is positive, but when it is not sequentially rational for $E$ to hire group 2 aggressively.  This implies that $p_E^*>p_2$ but $$(B-w)-w(1-p_2)(1-\phi_0)\geq 0.$$  This latter inequality is satisfied when \begin{equation}
\label{paretoG2}
p_2\geq \frac{w(1-\phi_0)}{B-w\phi_0}.
\end{equation}  
Thus, when $p_2\in \left[ \frac{w(1-\phi_0)}{B-w\phi_0}, p_E^*\right)$ banning the box Pareto dominates the box, leaving members of group 2 strictly better off; members of group 1 indifferent; and $E$ weakly better off.  When $p_2< \frac{w(1-\phi_0)}{B-w\phi_0}$ $E$ is strictly made worse off by the box, as $E$'s receives a negative payoff from hiring from group 2. 
\end{proof}

\bibliography{john}

\end{document}